\documentclass[fleqn,10pt]{wlscirep}
\usepackage[utf8]{inputenc}
\usepackage[T1]{fontenc}
\usepackage{lineno}
\usepackage{graphicx}
\usepackage{float}
\usepackage{subcaption}
\usepackage{adjustbox}
\usepackage{tabularx}
\usepackage{booktabs}

\title{ROCOv2: Radiology Objects in COntext Version 2, an Updated Multimodal Image Dataset}

\author[1]{Johannes R\"uckert}
\author[1,2,3,$\dagger$]{Louise Bloch}
\author[1,2,3,$\dagger$]{Raphael Br\"ungel}
\author[1,2,$\dagger$]{Ahmad Idrissi-Yaghir}
\author[1,4,$\dagger$]{Henning Sch\"afer}
\author[3,4,$\dagger$]{Cynthia S. Schmidt}
\author[3,5]{Sven Koitka} 
\author[1,2,3]{Obioma Pelka} 
\author[6]{Asma Ben Abacha}
\author[7]{Alba {G. Seco de Herrera}}
\author[8]{Henning M\"uller}
\author[4]{Peter A. Horn} 
\author[3,5]{Felix Nensa} 
\author[1,2,*]{Christoph M. Friedrich}

\affil[1]{Department of Computer Science, University of Applied Sciences and Arts Dortmund, Dortmund, Germany}
\affil[2]{Institute for Medical Informatics, Biometry and Epidemiology (IMIBE), University Hospital Essen, Essen, Germany}
\affil[3]{Institute for Artificial Intelligence in Medicine (IKIM), University Hospital Essen, Essen, Germany}
\affil[4]{Institute for Transfusion Medicine, University Hospital Essen, Essen, Germany}
\affil[5]{Institute of Diagnostic and Interventional Radiology and Neuroradiology, University Hospital Essen, Essen, Germany}
\affil[6]{Microsoft, Redmond, Washington, USA}
\affil[7]{University of Essex, Wivenhoe Park, Colchester, UK}
\affil[8]{University of Applied Sciences Western Switzerland (HES-SO), Switzerland}
\affil[*]{corresponding author(s): Christoph M. Friedrich (christoph.friedrich@fh-dortmund.de), technical inquiries: Johannes R\"uckert (johannes.rueckert@fh-dortmund.de)}
\affil[$\dagger$]{contributed equally}

\begin{abstract}
Automated medical image analysis systems often require large amounts of training data with high quality labels, which are difficult and time consuming to generate.
This paper introduces Radiology Object in COntext version 2 (ROCOv2), a multimodal dataset consisting of radiological images and associated medical concepts and captions extracted from the PMC Open Access subset.
It is an updated version of the ROCO dataset published in 2018, and adds 35,705 new images added to PMC since 2018. It further provides manually curated concepts for imaging modalities with additional anatomical and directional concepts for X-rays. The dataset consists of 79,789 images and has been used, with minor modifications, in the concept detection and caption prediction tasks of ImageCLEFmedical Caption 2023.
The dataset is suitable for training image annotation models based on image-caption pairs, or for multi-label image classification using Unified Medical Language System (UMLS) concepts provided with each image. In addition, it can serve for pre-training of medical domain models, and evaluation of deep learning models for multi-task learning.

\end{abstract}
\begin{document}

\flushbottom
\maketitle

\thispagestyle{empty}

\section*{Background \& Summary}

Recent years have seen tremendous progress in medical imaging.  The advent of deep learning techniques has enabled the development of sophisticated models for image analysis tasks. Multimodal image datasets play a crucial role in the development and validation of these models. One such dataset is Radiology Objects in COntext (ROCO)~\cite{ROCO2018}, which has enabled researchers to develop models for a wide range of tasks, including concept detection, caption generation, and image-text retrieval.

The first version of the ROCO dataset was introduced by Pelka et al.~\cite{ROCO2018} in 2018. It includes image-caption pairs from two classes: Radiology images from multiple imaging modalities and Out-of-Class images, such as synthetic radiology figures, digital art, and portraits, from peer-reviewed publications in the open-access subset of the biomedical literature database PubMed Central (PMC)~\cite{pmcoa}. 
The dataset contains 81,825 radiology images and 6127 out-of-class images. In addition to the images and their captions, the dataset provides keywords, Unified Medical Language System\textsuperscript{\textregistered} (UMLS\textsuperscript{\textregistered}) Semantic Types (SemTypes), and UMLS Concept Unique Identifiers (CUIs) for each image. This information makes the dataset suitable for training image annotation models based on image-caption pairs, or for multi-label image classification using the UMLS concepts provided with each image, e.g., to develop systems supporting structured medical reporting. The dataset is available on GitHub (available at \url{https://github.com/razorx89/roco-dataset}, accessed 2024-03-12) in the form of links to the publication and scripts to download and extract the images from them.

The ROCO dataset has been used in the medical caption tasks~\cite{ImageCLEFmedicalCaptionOverview2019,ImageCLEFmedicalCaptionOverview2020,ImageCLEFmedicalCaptionOverview2021,ImageCLEFmedicalCaptionOverview2022} at the Image Retrieval and Classification Lab of the Conference and Labs of the Evaluation Forum (ImageCLEF)~\cite{Müller2019}. 

ROCOv2 is the result of more than four years of updates and improvements to the original ROCO dataset. Due to the focus on radiological images, ROCOv2 does not include out-of-class images like ROCO, and to allow direct distribution of the images, only images from CC BY licensed articles (including CC BY-NC, but excluding CC BY-ND and CC BY-SA) are included.

Other changes include manually curated concepts, e.g., for modality of all images, anatomy and directionality of X-ray images, and improved concept extraction with the Medical Concept Annotation Toolkit v1.10.0 (MedCAT)~\cite{KRALJEVIC2021102083}, which is based on a newer version of the UMLS database and uses word embeddings instead of QuickUMLS~\cite{soldaini2016quickumls} that relies on direct dictionary matches. In addition, better concept filtering has been introduced.

The ROCOv2 dataset serves as a valuable resource for various applications and use cases in the medical domain, as it contains a vast amount of biomedical knowledge stored in the literature. One of the primary applications of this dataset is to train and evaluate models across different modalities for tasks such as image caption generation. By leveraging the multimodal nature of the dataset, researchers can develop models that accurately describe the content of radiological images, facilitating better understanding and communication of medical findings.
Furthermore, the ROCOv2 dataset can be utilized to build and train an efficient image retrieval system specifically tailored for the medical domain. Such a system would allow healthcare professionals to quickly search for relevant radiological images based on specific queries or similar case studies, enhancing the decision-making process and enabling more informed patient care. The image-caption pairs available in the dataset provide a rich foundation for training these retrieval models, ensuring accurate and relevant results. This can also be further extended to build a multimodal retrieval augmented generation (RAG) system that can be used in tasks such as generating detailed medical reports, or answering complex clinical questions.
Multimodal RAG also opens up additional possibilities beyond image retrieval. By combining the visual information from radiological images with the textual data from captions and associated medical literature, generative AI models can be trained or fine-tuned to produce more comprehensive and context-aware outputs. This approach can lead to the generation of synthetic data, which is particularly useful in cases where real-world medical data is scarce or difficult to obtain. The generated data can be used to augment existing datasets, improve model robustness, and support further research and development in the field of medical AI.

All sources of the dataset are openly available as part of the PMC Open Access Subset at the time of the publication of the dataset.

Summarizing the main contributions of this work:

\begin{itemize}
    \item Dataset of 79,789 radiological images with associated captions and medical concepts
    \item Possible use cases include training of image captioning, image retrieval and pre-training models
    \item Concepts were automatically generated from the captions, and combined with manually curated concepts for modality (all images), body region (X-ray only), and directionality (X-ray only)
    \item Images and captions were extracted from openly available publications with CC BY licenses in the PMC Open Access Subset
\end{itemize}

\section*{Related Work}
Since its initial release, the ROCO dataset has been used as the foundation for generating training and test data for multiple iterations of the ImageCLEFmedical Caption tasks, up to and including the most recent edition in 2023. Beyond these tasks, the ROCO dataset has proven to be a valuable resource for medical imaging research, resulting in its inclusion in several studies over the years. For instance, Eslami et al.~\cite{eslami-etal-2023-pubmedclip} investigated the effectiveness of Contrastive Language-Image Pre-training (CLIP)~\cite{RadfordKHRGASAM21} for the task of Medical Visual Question Answering (MedVQA) by leveraging the ROCO dataset to fine-tune CLIP for the medical domain. They chose the ROCO dataset for training due to its comprehensive collection, which includes various imaging modalities such as ultrasound, X-ray, PET, CT, MRI, and angiography from different human body regions, such as the head and pelvis. Their research resulted in the creation of PubMedCLIP, a specialized vision encoder that outperformed the general CLIP on two MedVQA benchmarks.

In addition to the ROCO dataset, several other datasets containing both image and text data have been published and used in medical imaging research.

One of the most notable examples is the MIMIC-CXR~\cite{Johnson2019} dataset. MIMIC-CXR is a large, publicly available collection of thorax radiology images paired with semi-structured free-text reports detailing radiological findings. The dataset includes 227,835 imaging studies from 65,379 patients, resulting in 377,110 images. Another dataset which also focuses on chest X-rays is the Open-I Indiana Chest X-ray collection~\cite{DemnerFushman2015}. The dataset consists of 3996 de-identified radiology reports and 8121 associated images. PADCHEST~\cite{BUSTOS2020101797} is a large dataset consisting of more than 160,000 chest X-ray images and associated reports from 67,000 patients. The reports are annotated with 174 different radiographic findings, 19 differential diagnoses, and 104 anatomical locations, hierarchically organized and mapped to standard UMLS~\cite{umls} terminology.
Additionally, manual annotations include bounding boxes for subfigures and their corresponding subcaptions for a subset of 2069 figures, resulting in 7507 subfigure-subcaption pairs. Compared to this work, MIMIC-CXR, Open-I Indiana Chest X-ray, and PADCHEST focuses only on chest X-rays, whereas ROCOv2 includes a wide range of anatomical regions, medical concepts, and modalities.

In another work, Subramanian et al.~\cite{subramanian-etal-2020-medicat} introduced MedICaT, a dataset of medical images, captions, and textual references. The dataset contains 217,060 images sourced from 131,410 open-access biomedical papers featuring captions and inline references for 74\% of the figures. Additionally, manual annotations including bounding boxes for sub-figures and their corresponding subcaptions, are provided for a subset of 2069 figures. Based on the dataset, the authors introduced the task of aligning subfigures with their corresponding subcaptions in compound figures and highlighted the valuable role of inline references in facilitating image-text matching. In comparison to the ROCOv2 dataset, MedICaT contains images distributed under the CC BY-ND and CC BY-SA licences which prohibit the distribution of processed images or require the transfer under the same license.

Recently, Lin et al.~\cite{lin2023pmcclip} proposed PMC-CLIP, a pre-trained model that uses biomedical documents for contrastive language-image pre-training based on PMC-OA, a biomedical dataset of 1.6 million image-caption pairs collected from the Open Access subset of PMC. The dataset covers various modalities and diseases, with the majority of the image annotation samples aligned at a fine-grained level, i.e., sub-figure and subcaption. The PMC-CLIP model achieves state-of-the-art results on several downstream tasks, including image-text retrieval on ROCO, MedMNIST~\cite{YangSN21} image classification, and medical VQA. To focus the PMC-OA dataset on biomedical images, filtering based on a keyword search and a deep learning-based classification model is used. Another pre-trained model, BiomedCLIP~\cite{zhang2024biomedclip}, performs well on various biomedical imaging tasks. It was pre-trained on PMC-15M, a large-scale dataset that includes a diverse range of 15 million biomedical image-text pairs. In contrast, ROCOv2 uses manual validation as a filtering step to achieve a high-quality radiological image dataset. Similar to the MedICaT dataset, PMC-OA contains images distributed under the CC BY-ND and CC BY-SA licenses, which are excluded from ROCOv2, so that the images of the dataset can be directly distributed.

Expanding on these datasets, the PMC-VQA~\cite{DBLP:journals/corr/abs-2305-10415} dataset has recently been introduced with a focus on the MedVQA task. The dataset contains 226,946 VQA pairs and 149,075 images, covering various medical modalities and diseases. It provides a comprehensive basis for the development and evaluation of MedVQA models. Data generation started with 381k image-caption pairs from the PMC-OA dataset. These captions were used with ChatGPT to generate five question-answer pairs per caption, which then underwent a filtering process. Experiments with models trained on the PMC-VQA dataset have demonstrated superior performance on established benchmarks such as the Visual Question Answering in Radiology (VQA-RAD)~\cite{Lau2018} and Semantically-Labeled Knowledge-Enhanced (SLAKE)~\cite{LiuZXMYW21} datasets. In addition, the authors proposed a manually verified test set that is more challenging and reflects the complexity of the real world. As the PMC-VQA dataset is based on the PMC-OA dataset, the same drawbacks apply, including no manual filtering and no filtering based on licences.

Leveraging the integration of visual and linguistic elements in medical datasets, Moor et al.~\cite{DBLP:journals/corr/abs-2307-15189} proposed Med-Flamingo, a multimodal few-shot learner adapted to the medical domain. It is based on OpenFlamingo-9B~\cite{DBLP:journals/corr/abs-2308-01390} and has been further pre-trained on paired and linked medical image-text data from publications and textbooks. Its unique strength lies in generative MedVQA, especially for open-ended questions similar to United States Medical Licensing Examination (USMLE) style problems. It has demonstrated its effectiveness by improving performance in generative MedVQA by up to 20\% on clinician ratings. The model was fine-tuned using the PMC-OA dataset. This dataset and the related problems have already been discussed.

In summary, while existing datasets such as MIMIC-CXR, PADCHEST, and PMC-OA have contributed to the field of medical imaging research, they have certain limitations. These datasets either focus on specific anatomical regions (e.g., chest X-rays), have license restrictions, or do not have manual validation. ROCOv2 aims to address some of these limitations by providing a diverse, manually validated dataset covering a wide range of anatomical regions, medical concepts, and modalities. In addition, by including only images with permissive licenses, ROCOv2 allows for the distribution of the dataset, facilitating its use in various research applications.

\section*{Methods}

\subsection*{Dataset Creation}

\begin{figure}[ht]
\centering
\includegraphics[width=\linewidth]{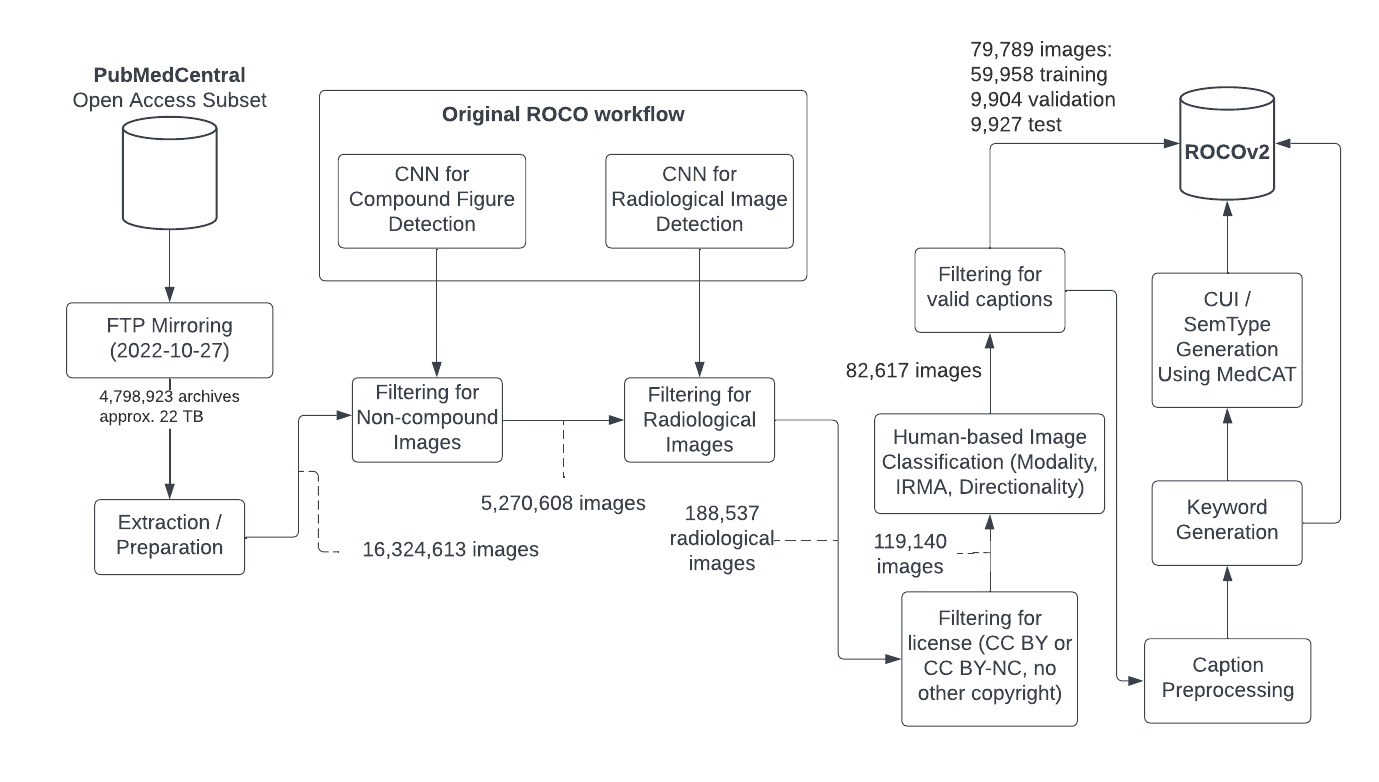}
\caption{Schematic overview of the dataset creation workflow. Based on Pelka et al.~\cite{ROCO2018}}
\label{fig:overview}
\end{figure}

Figure~\ref{fig:overview} shows a schematic overview of the dataset creation workflow described below.
The first step in creating the ROCOv2 dataset was to download the full PMC Open Access Subset via FTP, including all archives added until 2022-10-27. 4,798,923 archives, occupying 22 TB of disk space, were downloaded in this manner.

After extracting the archives, which include the PDF of the paper, as well as any images contained in the paper and an XML representation of the paper, more than 16,324,613 million extracted images are run through two binary classification models. The first is used to filter for non-compound images, while the second is used to filter for radiological images. The models are part of the original ROCO workflow described in~\cite{ROCO2018} and achieved accuracies of about 90\% and 98.6\%, respectively.

After this filtering step, 188,537 images (102,807 articles) were left, which were further reduced to 119,140 images (67,357 articles) by filtering out images from papers not licensed under CC BY or CC BY-NC as well as images which are subject to copyright of other commercial organizations or individuals, and by removing 2056 duplicates identified using AntiDupl v2.3.10 (available at \url{https://github.com/ermig1979/AntiDupl/}, accessed 2023-11-10).

From the remaining images, the new validation and test sets were created using 21,545 images from 2021 and 2022 that were not previously used in the ImageCLEFmedical Caption datasets. They were manually annotated for modality (angiography, CT, MRI, PET, ultrasound, X-ray, and combined modalities), Image Retrieval in Medical Applications (IRMA)~\cite{lehmann2003irma_classification} body region (X-ray only), and directionality (X-ray only) and combined with 34,900 images that were part of the original ROCO dataset and 28,086 images that had been used in previous ImageCLEFmedical Caption datasets. Stratified random sampling based on the manually curated concepts was used to divide the images into validation and test sets, and generated concepts that did not appear in the training set were removed from the validation and test sets.

The resulting 84,530 images (46,904 articles) were filtered for valid captions.
1528 images with non-English captions were removed. In addition, very short captions without relevant information (e.g., ``Figure 1'') were removed, resulting in a final dataset of 79,789 images (44,975 articles), with 59,958 images in the training set, 9904 images in the validation set, and 9927 images in the test set with 1947 unique CUIs overall, 1947 in the training set, 1760 in the validation set and 1754 in the test set. The detailed labeling and concept generation workflow is described in the next section. Compared to the dataset used in the ImageCLEFmedical Caption 2023 task, approximately 1500 non-radiological images were removed, further improving the quality of the dataset. 

Of the 81,825 radiology images in the original ROCO dataset, 33,645 were incorporated into ROCOv2~\cite{zenodo}, with the rest being excluded due to their license. They were combined with 46,144 images which have been used in ImageCLEFmedical Caption challenge datasets from 2021 to 2023.

\subsection*{Caption Processing and Concept Extraction}

To extract concepts from the captions, several pre-processing and filtering steps were performed. First, captions in languages other than English were excluded to focus the analysis on English-language concepts. This was done using the fastText~\cite{joulin-etal-2017-bag} language identification model. Captions identified as non-English with a confidence level greater than 45\% were excluded from the dataset. To reduce the risk of erroneously removing English captions, any caption identified as non-English with a confidence level of less than 45\% was retained under the assumption that it was likely written in English. Of all the captions, 1528 were identified as non-English. The bar chart in Table~S13 in the Supplementary Information shows the frequency of non-English captions across the different languages in the dataset. French captions were the most common with a count of 1413, followed by Portuguese and Spanish with 55 and 48 captions, respectively. Next, URLs within the captions were identified and removed, as they often do not provide relevant information for concept extraction. In addition, some captions were identified as consisting entirely of LaTeX code, and these were also removed from the dataset. Empty captions and those containing minimal information, such as ``xxx'', were discarded during pre-processing.

After the initial pre-processing, the remaining captions were further processed to extract relevant concepts using the Medical Concept Annotation Toolkit (MedCAT)~\cite{KRALJEVIC2021102083} framework. MedCAT is a robust tool specifically designed for extracting and linking biomedical concepts from unstructured text. It was trained on the MIMIC-III~\cite{Johnson2016} dataset (as of 2022-03-15) and links to Systematized Nomenclature of Medicine Clinical Terms (SNOMED CT) IDs. The SNOMED CT IDs were then mapped to UMLS2022AB release CUIs and semantic types (TUIs), which were then used for concept extraction and filtering.

During concept extraction, a frequency cutoff was implemented that retained only concepts that exceeded a frequency threshold of 10. In this way, low-frequency concepts, of which only a few examples were available, were effectively filtered out. By linking concepts to the UMLS, associated semantic types were filtered to focus on concepts that are likely to be visually observable and interpretable in the images. For example, concepts with the associated UMLS semantic type T029 (Body Location or Region) or T060 (Diagnostic Procedure) are relevant, while concepts of semantic type T054 (Social Behavior) cannot be derived from the image through a model. Specific concept filters were then manually applied to exclude UMLS concepts that could not be directly associated with the image content, such as temporal or qualitative aspects of certain concepts. Blacklisted concepts often contain qualifiers that would distract from the actual interest, and would also introduce bias since qualifiers are used in highly individual and variable ways by the original authors. Entity linking systems sometimes tend to incorrectly link concepts with ambiguous synonyms, e.g. C0994894 (Patch Dosage Form) may be linked if the caption refers to a region described as patchy. In the case of high frequency of such concepts, they have been manually mapped to the correct CUI.

In addition to the described automatic concept extraction, further manual creation and validation of basal concepts were performed. As in the original ROCO dataset~\cite{ROCO2018}, this mainly focused on the seven supported image modalities (angiography, combined modalities (e.g. PET/CT), CT, MRI, PET, ultrasound, and X-ray). However, for the ROCOv2 dataset, additional concepts were introduced for the X-ray modality, focusing on: (i) the displayed or described body region, and (ii) the directionality on which a projection was based. The body region classification was based on the IRMA classification~\cite{lehmann2003irma_classification}, which distinguishes eight different regions (abdomen, breast, chest, cranium, lower extremity, pelvis, spine, and upper extremity). The directionality classification was based on a reduced set of the most commonly used directions (coronal anteroposterior, coronal posteroanterior, sagittal, and transversal), but was introduced for experimental purposes only.

The reason for the manual creation of concepts is that in many cases the captions either do not explicitly provide information about them (e.g. "T2 weighted" implies MRI modality, "cholangiography" implies abdominal region).

The manual concept creation pipeline involved an initial manual classification of subsets of tens of thousands (modality, directionality) or several thousand (body region) images. This was done by two annotators for image modalities and X-ray body regions, and a single annotator for X-ray directionalities. The respective annotation guidelines followed by the annotators are provided in a distilled form as supplementary material. Deep learning image classification models were then trained on these subsets to pseudo-label the remaining images as a preliminary sorting method. These were then manually curated again to resolve errors in the classification models. Finally, the quality of the manually created concepts was validated by a radiologist on representative subsets for each category, with results described in the technical validation section.

\begin{figure}[ht]
    \centering
    \includegraphics[width=\textwidth]{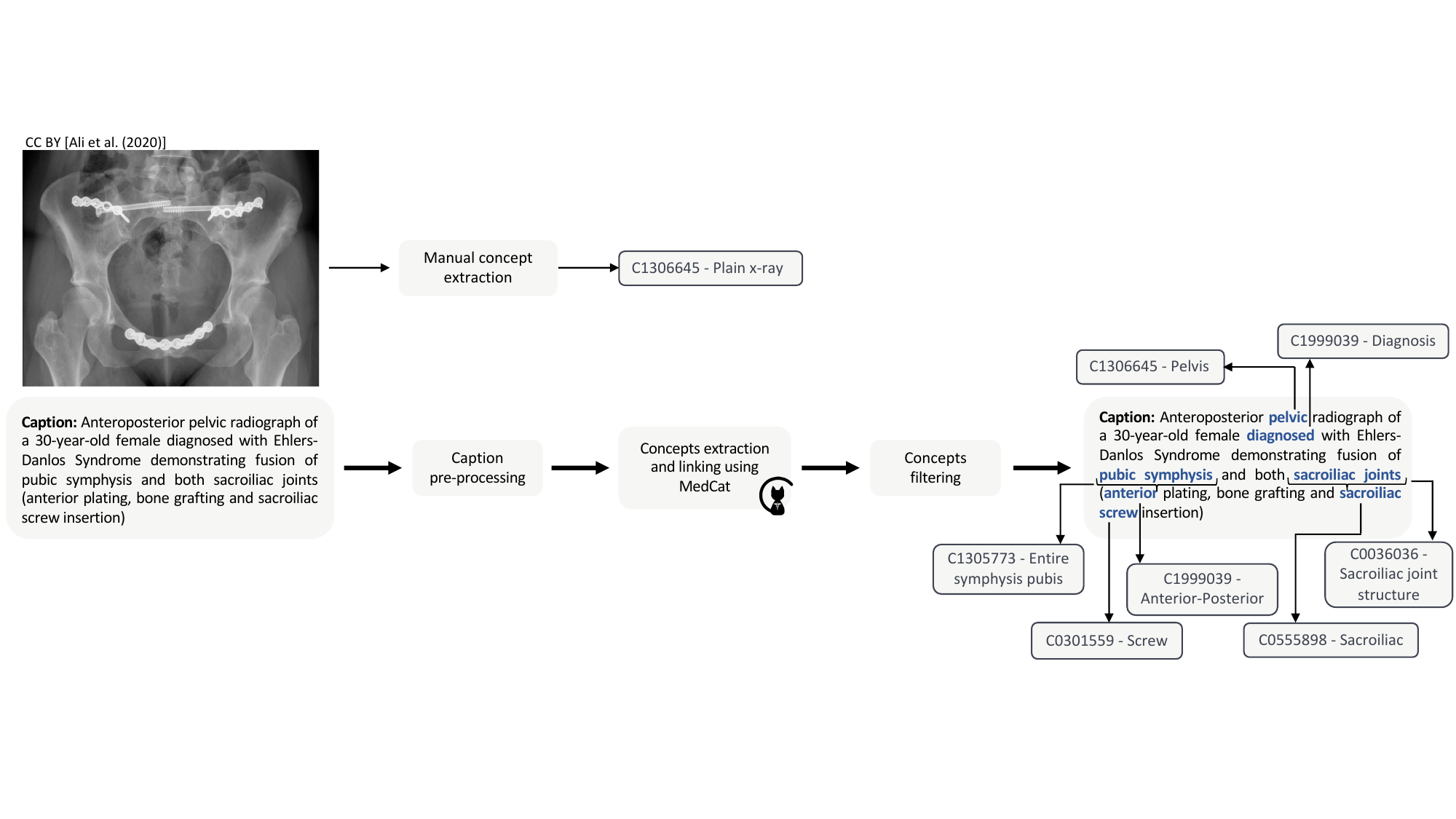}
    \caption{Concept extraction and caption pre-processing workflow. Radiology image taken from Ali et al.~\cite{jcm9123992}}
    \label{fig:concept-creation}
\end{figure}

The concepts from both manual and automatic extraction were combined, with priority given to the manually curated concepts. Automatically extracted modality concepts were included only for combined modalities (DRCO). The manually identified anatomy and directionality concepts of X-rays were not checked for conflicting concepts from automatic extraction to be integrated.
Figure~\ref{fig:concept-creation} shows the concept extraction and caption pre-processing workflow.

\section*{Data Records}

The ROCOv2 dataset files are available on Zenodo~\cite{zenodo}. It contains images, captions, and concepts for training, validation, and test splits, as well as image license information.

\begin{itemize}
  \item \textbf{``\{train,valid,test\}\_images.zip'':} JPEG images of various sizes taken directly from PMC with the filename format \texttt{ROCOv2\_2023\_XXXXXX.jpg}.
  \item \textbf{``\{train,valid,test\}\_captions.csv'':} Two-column comma-separated value (CSV) files with image filenames and corresponding captions (escaped with double quotes if necessary).
  \item \textbf{``\{train,valid,test\}\_concepts.csv'':} Two-column CSV files with image filenames and corresponding medical concept CUIs separated by semicolons. Includes manually curated and automatically generated concepts.
  \item \textbf{``\{train,valid,test\}\_concepts\_manual.csv'':} Two-column CSV files with image filenames and corresponding medical concept CUIs separated by semicolons. Includes manually curated concepts only.
  \item \textbf{``cui\_mapping.csv'':} Two-column CSV file with CUIs and their canonical name.
  \item \textbf{``license\_information.csv'':} Four-column CSV file with image filename, PMCID, CC BY attribution string, and PMC article link.
\end{itemize}

A dataset analysis is performed as part of the technical validation in the following section.

\section*{Technical Validation}

The ROCOv2~\cite{zenodo} dataset is based on the dataset used in the medical caption task~\cite{ImageCLEFmedicalCaptionOverview2023} at the ImageCLEF 2023~\cite{ImageCLEF2023}, where participants had access to the training and validation sets after signing a user agreement. ImageCLEF 2023 consists of the ImageCLEFmedical, ImageCLEFfusion, and ImageCLEFaware labs, where the ImageCLEFmedical lab is divided into the subtasks MEDIQA-Sum (natural language semantic retrieval), Caption, GANs (medical image generation) and MedVQA-GI (gastrointestinal visual question answering). The ImageCLEFmedical Caption task consists of two subtasks: concept detection and caption prediction.

All results are also described in detail in the overview paper~\cite{ImageCLEFmedicalCaptionOverview2023}.

Since several improvements were made to the dataset compared to the one used in ImageCLEFmedical Caption 2023, such as the removal of approximately 1500 non-radiological images and the addition of approximately 3000 missing manually curated concepts, baseline results are reported separately for both datasets.

The results for both subtasks show that the baseline models achieve similar results on the ImageCLEF dataset as the challenge participants while performing better on the ROCOv2~\cite{zenodo} dataset, showing the improved dataset quality. The baseline results, along with several years of competitive and improving scores for both subtasks in the context of the ImageCLEFmedical Caption challenges show the suitability of the dataset for training models for concept detection and caption prediction.

\subsection*{Concept Detection}

For the concept detection, participants are asked to predict a set of concepts defined by the UMLS CUIs~\cite{umls} based on the visual information provided by the radiology images, which can help in the development of systems supporting structured medical reporting. The balanced precision and recall trade-off were measured in terms of sample-averaged F1-scores, with a separate F1-score being calculated for manually curated concepts.

In the ImageCLEFmedical 2023 Caption challenge, the best team achieved an F1-score of 0.5223 using an ensemble of three multi-label classification models with different architectures~\cite{AUEB-NLP-Group2023}. Additional results are shown in Table~\ref{tb:results_concepts}.

\begin{table*}[ht!]
\centering
  \caption{Performance of the participating teams in the ImageCLEFmedical 2023 Concept Detection subtask. Only the best run based on the achieved sample-averaged F1-score is listed for each team, together with the corresponding secondary sample-averaged F1-score based on manual annotations as well as the team rankings based on the primary and secondary F1-score. The full results are shown in the overview paper~\cite{ImageCLEFmedicalCaptionOverview2023}. The best results are highlighted in bold. For comparison, the baseline results on the ImageCLEF dataset and on the ROCOv2~\cite{zenodo} dataset are included at the bottom.}
    \label{tb:results_concepts}
  \begin{tabular}{lrrrr}
\toprule
    \textbf{Group Name} & \textbf{F1} & \textbf{Secondary F1} & \textbf{Rank (secondary)} \\
\midrule
    AUEB-NLP-Group~\cite{AUEB-NLP-Group2023} & \textbf{0.5223} & 0.9258 & 1 (2) \\
    KDE-Lab\_Med~\cite{kdelab2023} & 0.5074 & \textbf{0.9321} & 2 (1) \\
    VCMI~\cite{vcmi2023} & 0.4998 & 0.9162 & 3 (3) \\
    IUST\_NLPLAB~\cite{IUST_NLPLAB2023} & 0.4959 & 0.8804 & 4 (6) \\
    Clef-CSE-GAN-Team~\cite{ClefCSEGAN2023} & 0.4957 & 0.9106 & 5 (4) \\
    CS\_Morgan~\cite{CSMorgan2023} & 0.4834 & 0.8902 & 6 (5) \\
    SSNSheerinKavitha~\cite{SSNMLRG2023} & 0.4649 & 0.8603 & 7 (7) \\
    closeAI2023~\cite{closeAI2023} & 0.0900 & 0.2152 & 8 (8) \\
    SSN\_MLRG~\cite{SSNMLRG2023} & 0.0173 & 0.1122 & 9 (9) \\
\bottomrule
    Baseline EfficientNetB0 ImageCLEF & 0.5099 & 0.9309 & – \\
    Baseline EfficientNetB0 ROCOv2 & 0.5811 & 0.9353 & – \\
    Baseline EfficientNetv2-s ImageCLEF & 0.5215 & 0.9407 & – \\
    Baseline EfficientNetv2-s ROCOv2 & 0.5925 & 0.9430 & – \\
\bottomrule
\end{tabular}
\end{table*}

As previously mentioned, the ROCOv2~\cite{zenodo} dataset is an improved version of the dataset provided in the ImageCLEFmedical Caption task. In order to compare the challenge results to the results of the ROCOv2 dataset, two baseline models, namely an EfficientNet-B0~\cite{Tan2019EfficientNet} and an EfficientNetv2-s~\cite{Tan2021EfficientNetv2}, were additionally trained and results for both datasets are given.

The implementation was developed using PyTorch v2.0.1~\cite{Paszke2019PyTorch}, and all experiments were run on an NVIDIA\textsuperscript{\textregistered} DGX-1 (available at \url{https://www.nvidia.com/en-gb/data-center/dgx-1/}, accessed 2023-11-10) supercomputer, with NVIDIA\textsuperscript{\textregistered} V100 (available at \url{https://www.nvidia.com/en-us/data-center/v100/}, accessed 2023-11-10) Graphical Processing Units (GPUs) containing 16 GB of memory. The execution environment was an NVIDIA\textsuperscript{\textregistered}-optimized (available at \url{https://github.com/NVIDIA/nvidia-docker}, accessed 2023-11-10) Docker~\cite{merkel2014docker} container, running a Deepo (available at \url{https://github.com/ufoym/deepo}, accessed 2023-11-10) image. All experiments were executed using a single GPU.

For both models, a grid search was performed on the validation dataset for hyperparameter tuning to identify the best combination of optimizer and learning rate. The values [1e-1,1e-2,...,1e-5] are used as candidates for the learning rate. In addition, the Adam~\cite{Kingma2015Adam}, Stochastic Gradient Descent (SGD), and Root Mean Square Propagation (RMSProp) optimizers were tested. After hyperparameter tuning, the final models were trained on the entire training and validation dataset. To train both models, the training augmentation pipeline includes loading the images with an image size of 1.25 times the model image size, random horizontal and vertical flipping with a probability of 0.5 each, random cropping to the image size of the model, and image normalization. The validation and test augmentation pipelines include loading the images with an image size of 1.25 times the model image size, center cropping to the image size of the model and image normalization.
The loss function used is a multi-label soft margin loss. A sigmoid activation function is used for all model outputs with a threshold of 0.5. All models are trained using mixed precision~\cite{Micikevicius2017MixedPrecision} for 20 epochs. For the remaining hyperparameters, the default values in PyTorch are used.

The model based on the EfficientNet-B0 architecture was pre-trained on the ImageNet-1k dataset~\cite{Deng2009ImageNet}. This model was trained with a batch size of 256, a drop rate of 0.2, and an image size of 224. During hyperparameter tuning, the Adam optimizer, trained with a learning rate of 1e-3, achieved the best sample-averaged F1-score on the validation dataset for both datasets (ImageCLEFmedical Caption and ROCOv2~\cite{zenodo}). The final model achieved a sample-averaged F1-score of 0.5099 (secondary sample-averaged F1-score: 0.9309) for the ImageCLEFmedical Caption test set and a sample-averaged F1-score of 0.5811 (secondary sample-averaged F1-score: 0.9353) for the ROCOv2 test set. The EfficientNetv2-s model was pre-trained on the ImageNet-21k dataset~\cite{RidnikBNZ21}. This model was trained with a batch size of 92, a drop rate of 0.2, and an image size of 300. During hyperparameter tuning, the RMSProp optimizer trained with a learning rate of 1e-4 achieved the best F1-score on the validation dataset for both datasets (ImageCLEFmedical Caption and ROCOv2~\cite{zenodo}). The final model achieved a sample-averaged F1-score of 0.5215 (secondary sample-averaged F1-score: 0.9407) for the ImageCLEFmedical Caption test set and a sample-averaged F1-score of 0.5925 (secondary sample-averaged F1-score: 0.9430) for the ROCOv2 test set.

\subsection*{Caption Prediction}

The caption prediction aims to automatically generate captions for the radiology images provided. In ImageCLEFmedical Caption, the performance of caption prediction is evaluated based on BERTScore~\cite{zhangBERTScoreEvaluatingText2020}, which is a metric that computes a similarity score for each token in the generated text with each token in the reference text. Several other metrics were also calculated and published, to illustrate how difficult the evaluation of caption similarity is: First, the Recall-Oriented Understudy for Gisting Evaluation (ROUGE)~\cite{linROUGEPackageAutomatic2004} score was adopted as a secondary metric that counts the number of overlapping units such as n-grams, word sequences, and word pairs between the generated text and the reference. In addition to ROUGE, the Metric for Evaluation of Translation with Explicit ORdering (METEOR)~\cite{denkowskiMeteorUniversalLanguage2014} was explored, which is a metric that evaluates the generated text by aligning it to reference and calculating a sentence-level similarity score. Furthermore, the Consensus-based Image Description Evaluation (CIDEr)~\cite{vedantamCIDErConsensusbasedImage2015} metric was also adopted. CIDEr is an automatic evaluation metric that calculates the weights of n-grams in the generated text, and the reference text based on term frequency and inverse document frequency (TF-IDF) and then compares them based on cosine similarity.
Another metric used is the BiLingual Evaluation Understudy (BLEU) score~\cite{bleu2002}, which is a geometric mean of n-gram scores from 1 to 4. For this task, the focus was on the BLEU-1 score, which takes into account unigram precision.
Bilingual Evaluation Understudy with Representations from Transformers (BLEURT)~\cite{sellam-etal-2020-bleurt} is specifically designed to evaluate natural language generation in English. It uses a pre-trained model that has been fine-tuned to emulate human judgments about the quality of the generated text.
CLIPScore~\cite{hessel-etal-2021-clipscore} is an innovative metric that diverges from the traditional reference-based evaluations of image captions. Instead, it aligns with the human approach of evaluating caption quality without references by evaluating the alignment between text and image content.

For the caption prediction subtask at ImageCLEFmedical 2023, the best team achieved a BERTScore of 0.6413 with an encoder-decoder framework with subsequent reinforcement learning~\cite{CSIRO2023}. Additional results are shown in Tables~\ref{tb:results_caption} and \ref{tb:results_caption_additional}.

\begin{table*}[h!]
\centering
  \caption{Performance of the participating teams in the ImageCLEFmedical 2023 caption prediction subtask. Only the best run based on the achieved BERTScore is listed for each team, together with the corresponding secondary ROUGE score as well as the team rankings based on the primary BERTScore and secondary ROUGE score. Additional scores are shown in Table~\ref{tb:results_caption_additional}. The full results are shown in the overview paper~\cite{ImageCLEFmedicalCaptionOverview2023}. The best results are highlighted in bold. For comparison, the baseline results on the ImageCLEF dataset and on the ROCOv2 dataset are included at the bottom.}
    \label{tb:results_caption}
  \begin{tabular}{lrrrr}
\toprule
    \textbf{Group Name} & \textbf{BERTScore} & \textbf{ROUGE} & \textbf{Rank (secondary)} \\
\midrule
    CSIRO~\cite{CSIRO2023} & \textbf{0.6413} & 0.2463 & 1 (3) \\
    closeAI2023~\cite{closeAI2023} & 0.6281 & 0.2401 & 2 (4) \\
    AUEB-NLP-Group~\cite{AUEB-NLP-Group2023} & 0.6170 & 0.2130 & 3 (8) \\
    PCLmed~\cite{PCLmed2023} & 0.6152 & 0.2528 & 4 (2) \\
    VCMI~\cite{vcmi2023} & 0.6147 & 0.2175 & 5 (7) \\
    KDE-Lab\_Med~\cite{kdelab2023} & 0.6145 & 0.2223 & 6 (5) \\
    SSN\_MLRG~\cite{SSNMLRG2023} & 0.6019 & 0.2112 & 7 (9) \\
    DLNU\_CCSE & 0.6005 & 0.2029 & 8 (10) \\
    CS\_Morgan~\cite{CSMorgan2023} & 0.5819 & 0.1564 & 9 (11) \\
    Clef-CSE-GAN-Team~\cite{ClefCSEGAN2023} & 0.5816 & 0.2181 & 10 (6) \\
    Bluefield-2023~\cite{Bluefield2023}  & 0.5780 & 0.1534 & 11 (12) \\
    IUST\_NLPLAB~\cite{IUST_NLPLAB2023} & 0.5669 & \textbf{0.2898} & 12 (1) \\
    SSNSheerinKavitha~\cite{SSNMLRG2023} & 0.5441 & 0.0866 & 13 (13) \\
\bottomrule
    Baseline ImageCLEF & 0.6217 & 0.2318 & – \\
    Baseline ROCOv2 & 0.6264 & 0.2352 & – \\
\bottomrule
\end{tabular}
\end{table*}
\begin{table*}[h!]
\centering
  \caption{Performance of the participating teams in the ImageCLEFmedical 2023 caption prediction subtask for additional metrics BLEURT, BLEU, METEOR, CIDEr and CLIPScore. These correspond to the best BERTScore-based runs of each team, listed in Table~\ref{tb:results_caption}. The full results are shown in the overview paper~\cite{ImageCLEFmedicalCaptionOverview2023}. The best results are highlighted in bold. For comparison, the baseline results on the ImageCLEF dataset and on the ROCOv2 dataset are included at the bottom.}
    \label{tb:results_caption_additional}
  \begin{tabular}{lrrrrrr}
\toprule
    \textbf{Group Name} & \textbf{BLEURT} &  \textbf{BLEU} & \textbf{METEOR} & \textbf{CIDEr} & \textbf{CLIPScore} \\
\midrule
    CSIRO~\cite{CSIRO2023} & 0.3137 & 0.1615 & 0.0798 & 0.2025 & \textbf{0.8147} \\
    closeAI2023~\cite{closeAI2023} & \textbf{0.3209} & 0.1846 & 0.0873 & \textbf{0.2377} & 0.8075 \\
    AUEB-NLP-Group~\cite{AUEB-NLP-Group2023} & 0.2950 & 0.1692 & 0.0720 & 0.1466 & 0.8039 \\
    PCLmed~\cite{PCLmed2023} & 0.3166 & 0.2172 & 0.0921 & 0.2315 & 0.8021 \\
    VCMI~\cite{vcmi2023} & 0.3084 & 0.1653 & 0.0734 & 0.1720 & 0.8082 \\
    KDE-Lab\_Med~\cite{kdelab2023} & 0.3014 & 0.1565 & 0.0724 & 0.1819 & 0.8062 \\
    SSN\_MLRG~\cite{SSNMLRG2023} & 0.2774 & 0.1418 & 0.0615 & 0.1284 & 0.7759 \\
    DLNU\_CCSE & 0.2630 & 0.1059 & 0.0557 & 0.1332 & 0.7725 \\
    CS\_Morgan~\cite{CSMorgan2023} & 0.2242 & 0.0566 & 0.0436 & 0.0840 & 0.7593 \\
    Clef-CSE-GAN-Team~\cite{ClefCSEGAN2023} & 0.2690 & 0.1450 & 0.0702 & 0.1737 & 0.7893 \\
    Bluefield-2023~\cite{Bluefield2023} & 0.2716 & 0.1543 & 0.0601 & 0.1009 & 0.7837 \\
    IUST\_NLPLAB~\cite{IUST_NLPLAB2023} & 0.2230 & \textbf{0.2685} & \textbf{0.1004} & 0.1773 & 0.8068 \\
    SSNSheerinKavitha~\cite{SSNMLRG2023} & 0.2152 & 0.0749 & 0.0258 & 0.0143 & 0.6873 \\
\bottomrule
    Baseline ImageCLEF & 0.3093 & 0.1821 & 0.0813 & 0.1968 & 0.8172 \\
    Baseline ROCOv2 & 0.3113 & 0.1772 & 0.0821 & 0.2026 & 0.8207 \\
\bottomrule
\end{tabular}
\end{table*}

As a baseline for the caption prediction task, a model leveraging a vision encoder-decoder architecture~\cite{VinyalsTBE15} was employed. The encoder component was instantiated with the base-sized (available at \url{https://huggingface.co/google/vit-base-patch16-224-in21k}, accessed 2023-11-10) Vision Transformer (ViT) model~\cite{DosovitskiyB0WZ21}. This model is based on the transformer architecture and was pre-trained on the ImageNet-21k dataset~\cite{RidnikBNZ21} at resolution 224x224. To initialize the decoder, the BioMedLM (available at \url{https://huggingface.co/stanford-crfm/BioMedLM}, accessed 2023-11-10) model was used, which is a language model based on the GPT2 architecture~\cite{Radford2019LanguageMA}. This decoder-only transformer-based model has 2.7 billion parameters and a maximum context length of 1024 tokens. The BioMedLM training data is derived from the PubMed Abstracts and PMC sections of the Pile dataset, which contains approximately 50 billion tokens from 16 million abstracts and 5 million full-text articles from the biomedical literature. This training corpus provides BioMedLM with a strong understanding of biomedical language, making it uniquely suited for tasks in the biomedical domain. The model was trained using the Huggingface Transformers library for two epochs with a batch size of two and a gradient accumulation of two. The maximum input sequence length was set to 128 tokens. As an optimizer, the Adafactor optimizer was chosen due to its efficiency in memory usage and its adaptability in adjusting learning rates. In addition, fp16 mixed precision was used during training.

The baseline model was first trained and evaluated on the ImageCLEFmedical Caption dataset. During this initial evaluation, the model demonstrated good performance, achieving a BERTScore of 0.6217 and a ROUGE score of 0.2318. It also obtained a CLIP score of 0.8172, a BLEURT score of 0.3093, a BLEU score of 0.1821, a METEOR score of 0.0813, and a CIDEr score of 0.1968. The same model architecture was then trained and evaluated on the ROCOv2~\cite{zenodo} dataset. The model achieved a BERTScore of 0.6241 and a ROUGE score of 0.2325, along with a CLIP score of 0.8212, a BLEURT score of 0.3131, a BLEU score of 0.1836, a METEOR score of 0.0830, and a CIDEr score of 0.2026.

\subsection*{Dataset Analysis}

Table~\ref{table:desc_stats} shows the descriptive statistics derived from the dataset. On average, each caption contains about 21 words but can range from one word to 848 words. Each article contains about 1.76 image-caption pairs on average, with some articles having up to 28 image-caption pairs. In addition, on average, captions are annotated with about 3.36 UMLS concepts, with some captions having as many as 28 concepts.

\begin{table}[ht]
    \centering
    \caption{Descriptive statistics of the ROCOv2 dataset.}
    \label{table:desc_stats}
    \begin{tabular}{lrrr}
    \toprule
                                \textbf{Statistic} &   \textbf{Average} &  \textbf{Maximum} &  \textbf{Minimum} \\
    \midrule
                Caption Length (in words) & 20.91 &      778 &        1 \\
                     Captions per Article &  1.77 &       28 &        1 \\
    No. of Extracted Concepts per Caption &  3.36 &       28 &        1 \\
    \bottomrule
    \end{tabular}
\end{table}

\begin{figure}[ht]
    \centering
    \includegraphics[scale=0.7]{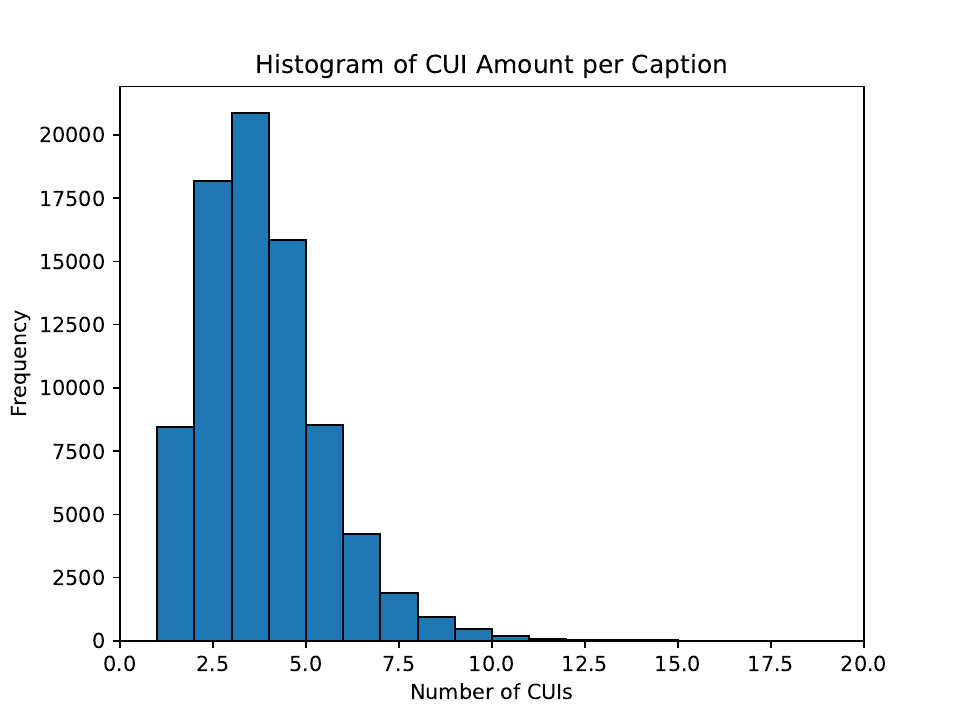}
    \caption{Histogram of CUI amount per caption.}
    \label{fig:freq_histo}
\end{figure}

The histogram in Figure~\ref{fig:freq_histo} shows the distribution of the number of concepts per image-caption pair. All captions have at least one concept.

Table~\ref{tab:umls_concepts} lists the ten most common UMLS concepts found in the dataset, excluding modality-related concepts that were set manually.

\begin{table}[ht]
    \caption{The ten most common UMLS concepts.}
    \centering
    \begin{tabular}{clr}
        \toprule
        \textbf{CUI} & \textbf{UMLS Term} & \textbf{Images} \\
        \midrule
        C0444611 & Fluid & 1779 \\
        C0027651 & Neoplasms & 1705 \\
        C0205207 & Cystic & 1469 \\
        C0018787 & Heart & 1288 \\
        C0023884 & Liver & 1244 \\
        C0012359 & Pathological Dilatation & 1084 \\
        C1266909 & Entire bony skeleton & 1077 \\
        C0006104 & Brain & 1058 \\
        C0032227 & Pleural effusion disorder & 1049 \\
        C0028259 & Nodule & 1032 \\
        \bottomrule
    \end{tabular}
    \label{tab:umls_concepts}
\end{table}

The ten most common semantic types (TUIs) are outlined in Table~\ref{tab:tui_freq}. These types provide insights into various medical concepts, ranging from diagnostic procedures to neoplastic processes.

\begin{table}[H]
\caption{The ten most common semantic types (TUIs). Note that each caption can contain several CUIs of the same semantic type.}
\centering
\begin{tabular}{llr}
\toprule
\textbf{TUI} & \textbf{Definition} &  \textbf{Frequency} \\
\midrule
    T023 & Body Part, Organ, or Organ Component & 80,723  \\
    T060 & Diagnostic Procedure & 79,750  \\
    T029 & Body Location or Region & 25,504 \\
    T082 & Spatial Concept & 16,801 \\
    T046 & Pathologic Function & 14,008 \\
    T047 & Disease or Syndrome & 12,311 \\
    T080 & Qualitative Concept & 5657 \\
    T030 & Body Space or Junction & 5078 \\
    T074 & Medical Device & 5074 \\
    T191 & Neoplastic Process & 4732 \\
    \bottomrule
\end{tabular}
    \label{tab:tui_freq}
\end{table}

\subsection*{Annotator-Radiologist Evaluation of Manual Concepts}
\label{sec:agreement}

To assess the overall quality and validity of the manually created concepts, an evaluation of the agreement between annotators and a radiologist was performed. To determine this, a representative subset of the dataset was created for each manual concept category: (i) modality of all images, (ii) displayed body region of X-ray modality images, and (iii) directionality of X-ray modality images. Representativeness was ensured by stratifying the corresponding labels. A radiologist then manually labeled each subset, independently going through the same process as the annotators and following the same labeling guidelines for each category.

Subsets were extracted from an internal, raw version of the original ImageCLEFmedical Caption 2023 dataset that included additionally the labels \texttt{OTHER} and \texttt{UNKNOWN} for later filtering and refinement purposes (e.g., out-of-class images, mixed-class images, uncertainty regarding distinct label assignment). Their exact meaning for each category is described in the annotation guidelines in the supplementary material. This was done to not bias the evaluation by applying premature filtering that may have excluded images that would have been valid in a radiologist's eye.

\begin{figure}[H]
    \centering
    \includegraphics[scale=0.3]{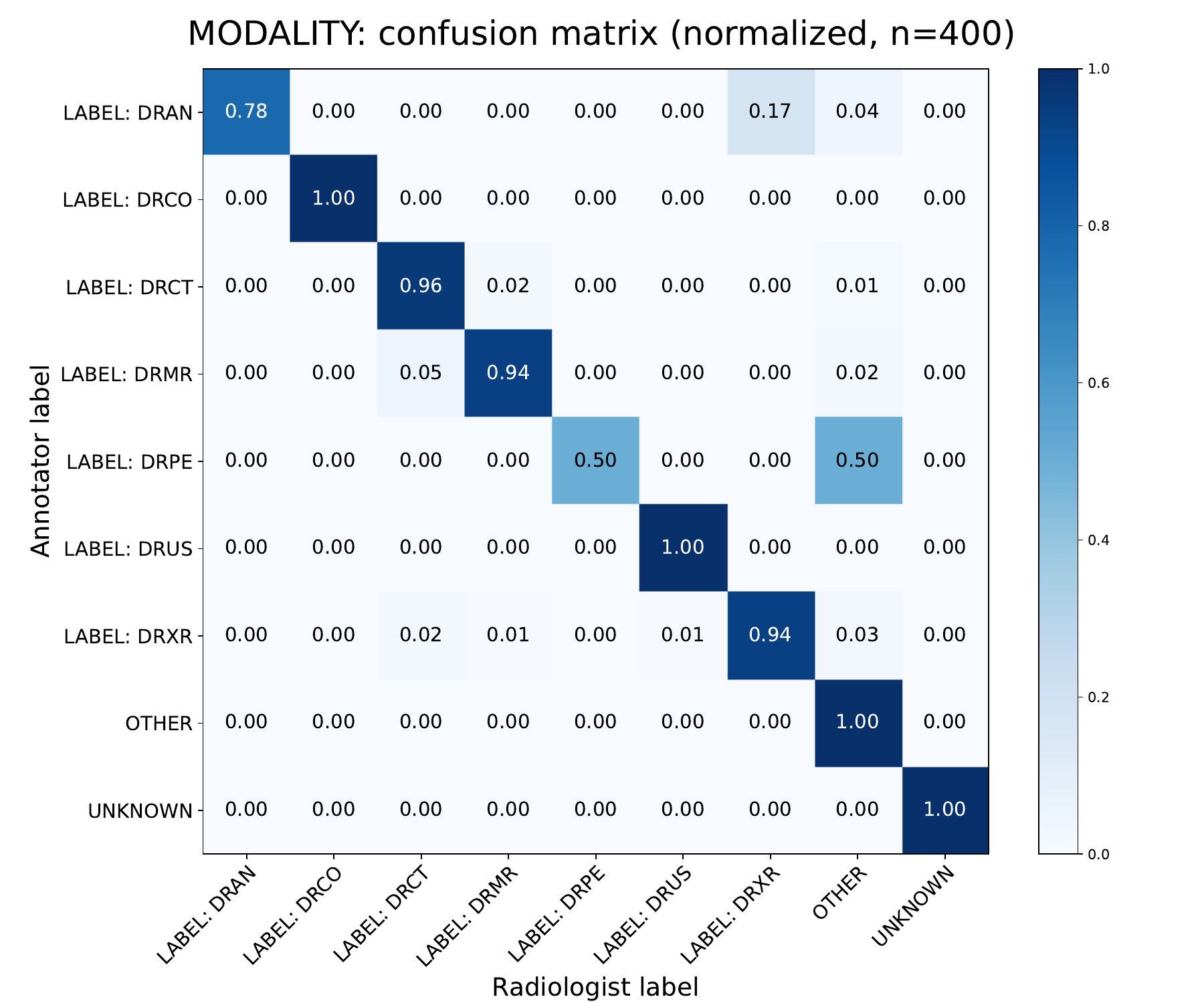}
    \caption{Annotator-Physician evaluation on manually labeled image modality for a representative subset of $n=400$ samples. Confusion matrix (normalized).}
    \label{fig:modality_cm_norm}
\end{figure}

\begin{figure}[H]
    \centering
    \includegraphics[scale=0.3]{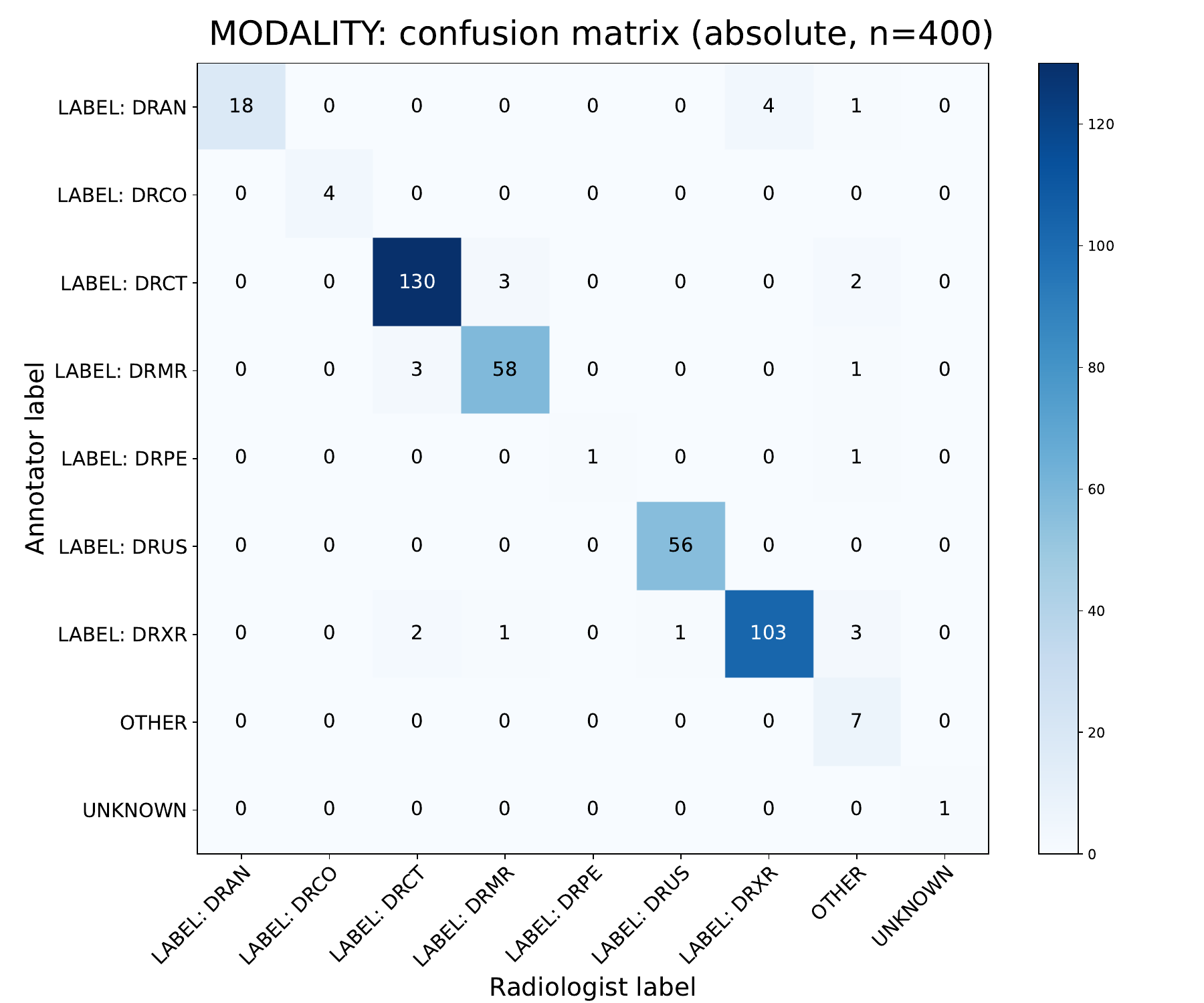}
    \caption{Annotator-Physician evaluation on manually labeled image modality for a representative subset of $n=400$ samples. Confusion matrix (absolute).}
    \label{fig:modality_cm_abs}
\end{figure}

\begin{figure}[H]
    \centering
    \includegraphics[scale=0.3]{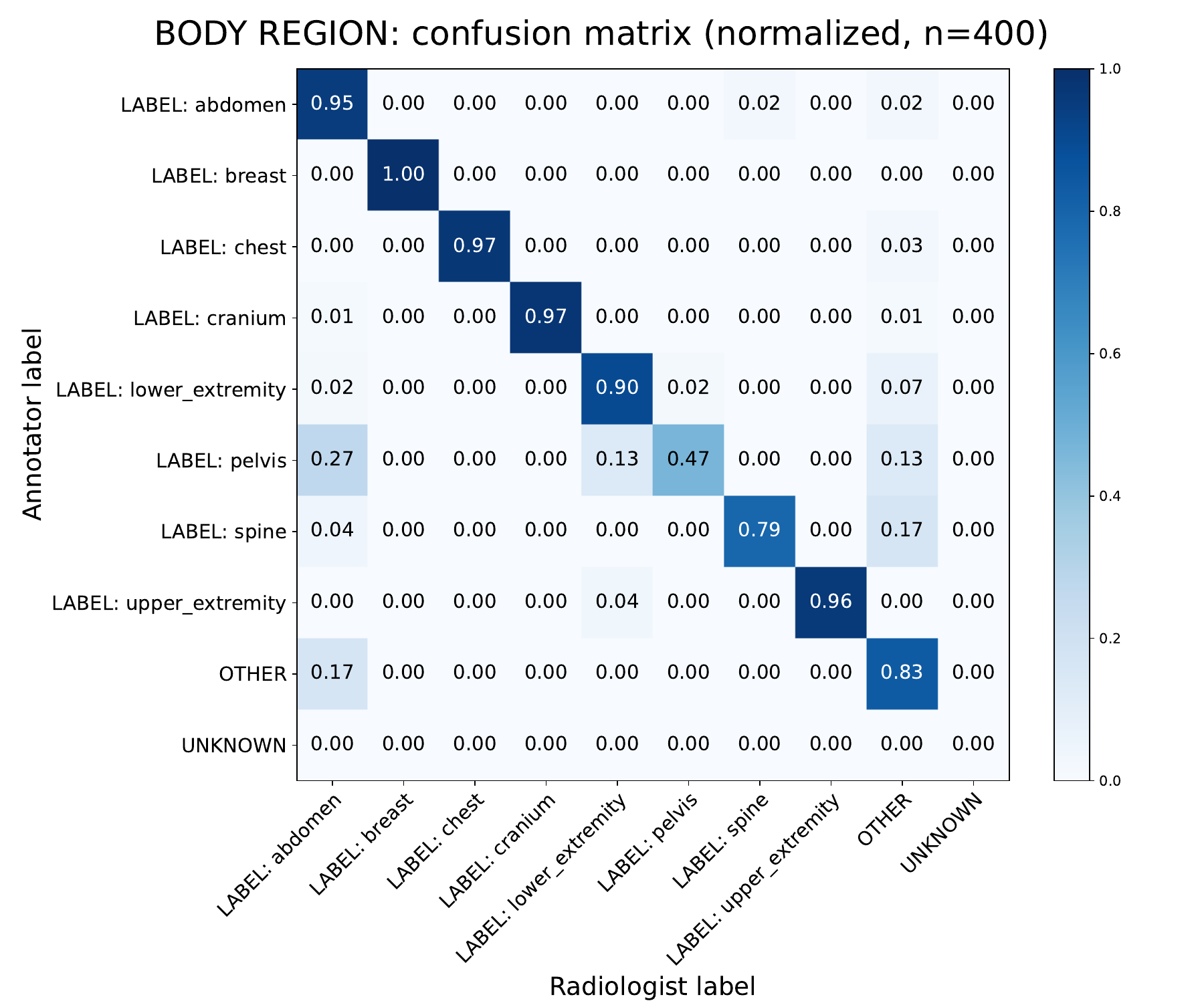}
    \caption{Annotator-Physician evaluation on manually labeled body regions in X-ray modality images for a representative subset of $n=400$ samples. Confusion matrix (normalized).}
    \label{fig:irma_cm_norm}
\end{figure}

\begin{figure}[H]
    \centering
    \includegraphics[scale=0.3]{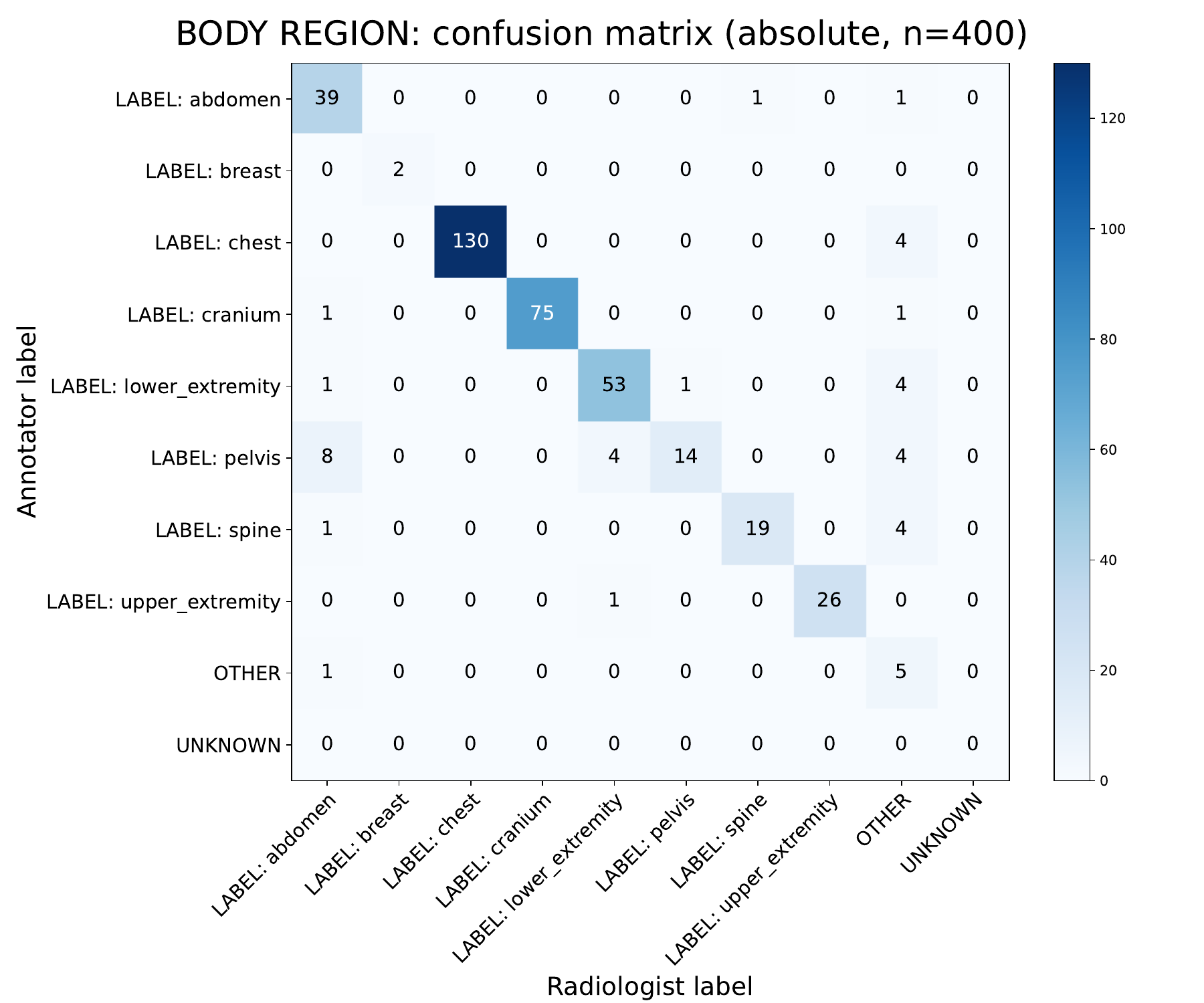}
    \caption{Annotator-Physician evaluation on manually labeled body regions in X-ray modality images for a representative subset of $n=400$ samples. Confusion matrix (absolute).}
    \label{fig:irma_cm_abs}
\end{figure}

\begin{figure}[H]
    \centering
    \includegraphics[scale=0.3]{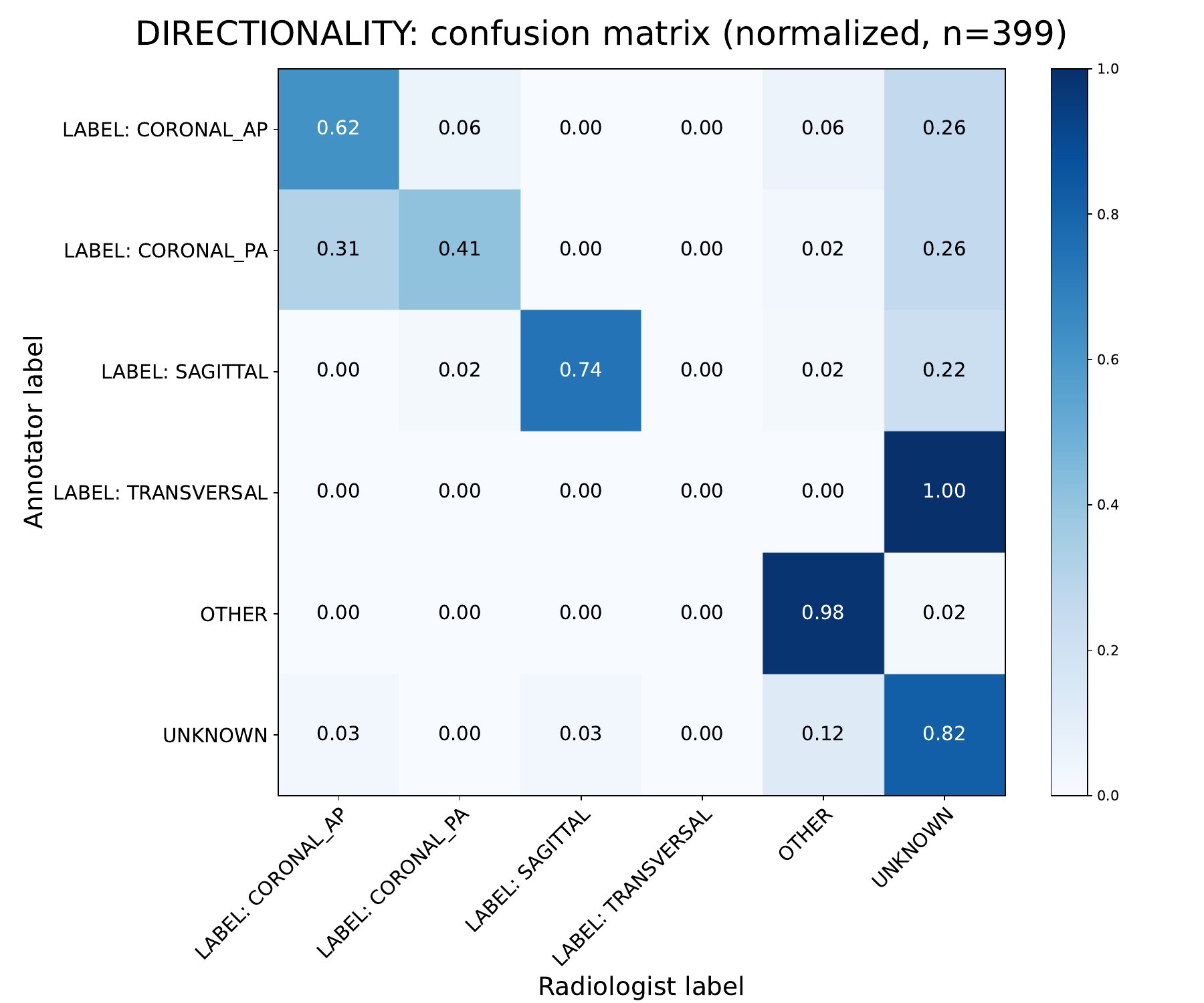}
    \caption{Annotator-Physician evaluation on manually labeled directionalities in X-ray modality images for a representative subset of $n=399$ samples. Confusion matrix (normalized).}
    \label{fig:directionality_cm_norm}
\end{figure}

\begin{figure}[H]
    \centering
    \includegraphics[scale=0.3]{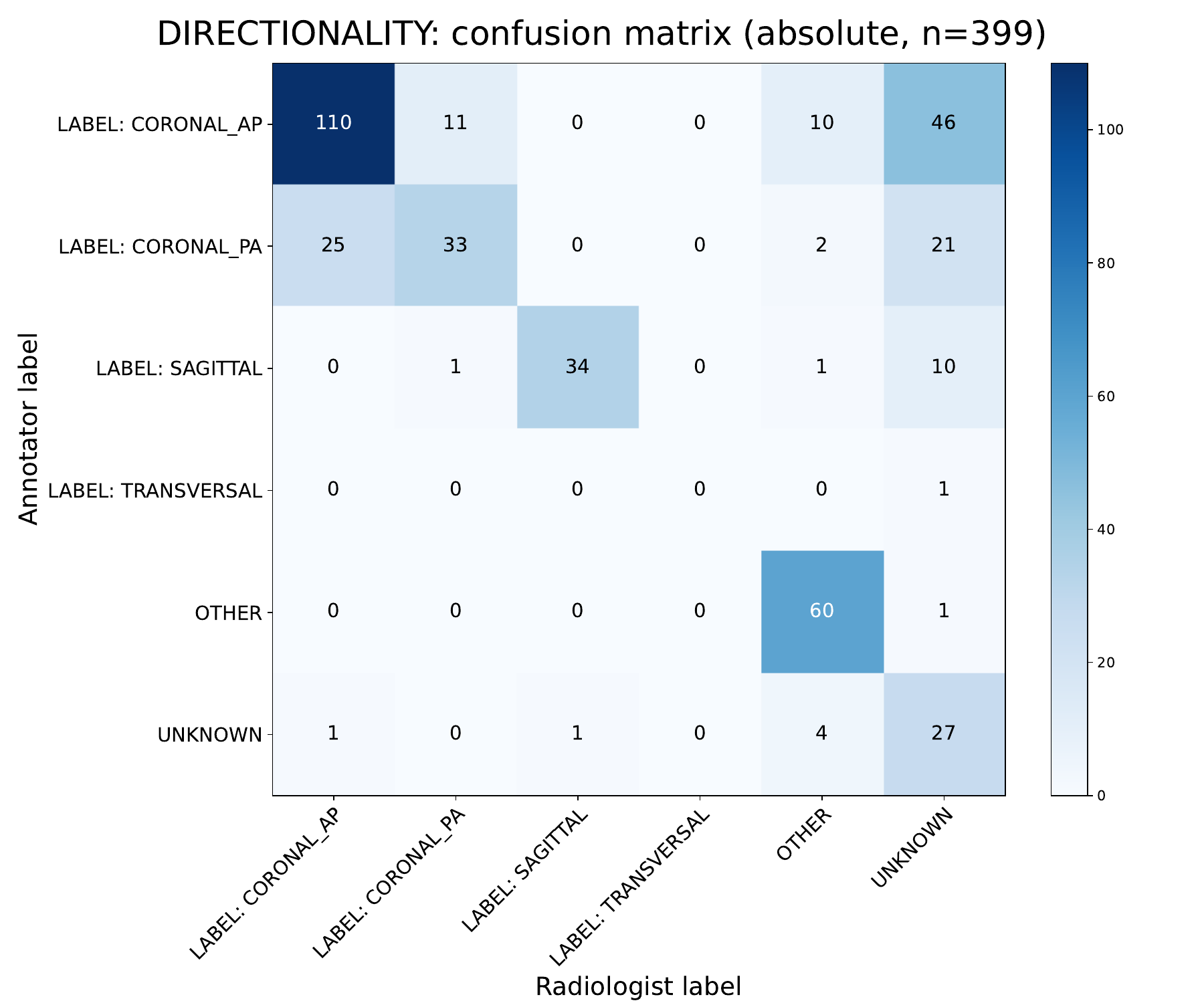}
    \caption{Annotator-Physician evaluation on manually labeled directionalities in X-ray modality images for a representative subset of $n=399$ samples. Confusion matrix (absolute).}
    \label{fig:directionality_cm_abs}
\end{figure}

To quantify inter-annotator agreement, confusion matrices based on annotator and radiologist labels were created and Cohen's $\kappa$~\cite{cohen1960kappa} analyses were performed. Corresponding normalized and absolute confusion matrices for modalities, body regions, and directionalities are shown in Figures~\ref{fig:modality_cm_norm}, \ref{fig:modality_cm_abs}, \ref{fig:irma_cm_norm}, \ref{fig:irma_cm_abs}, \ref{fig:directionality_cm_norm}  and \ref{fig:directionality_cm_abs}. The results of Cohen's $\kappa$ analysis are presented in Table~\ref{tab:cohens_kappa}.

A Cohen's $\kappa$ within an interval of [0.81, 1.00] can be interpreted as almost perfect, and within a range of [0.41, 0.60] as moderate agreement~\cite{landis1977kappa}. Thus, high values of $\kappa=0.928$ for image modalities and $\kappa=0.886$ for body regions indicate trustworthiness of manually curated concepts for both categories. However, a moderate value of $\kappa=0.557$ for directionalities highlights their experimental character. Identified reasons for decreased agreement are outlined in the Limitations section.

\begin{table}[htbp]
    \centering
    \caption{Overview of an evaluation on annotator-physician agreement for manually annotated concepts. Cohen's $\kappa$ with standard deviation (sd), the 95\% confidence interval (CIs) as well as observed agreements and those expected by chance are reported.}
    \label{tab:cohens_kappa}

    \begin{tabular}{lrrrr}
        \cmidrule{2-5}
        & \textbf{Cohen's $\kappa$ (sd)} & \textbf{$95$\% CI} & \textbf{Agreements (\%)} & \textbf{Expected by chance (\%)} \\
        \midrule
        Modality & 0.928 (0.015) & [0.899, 0.957] & 378 (94.50\%) & 93.9 (23.48\%) \\
        Body region$^\text{a}$ & 0.886 (0.017) & [0.851, 0.920] & 363 (90.75\%) & 76.2 (19.05\%) \\
        Directionality$^\text{a}$ & 0.557 (0.030) & [0.498, 0.616] & 264 (66.17\%) & 94.0 (23.57\%) \\
        \bottomrule
    \end{tabular}

    \smallskip
    \raggedright $^\text{a}$ Accounts solely for the X-ray modality. All other modalities did not receive manually labeled concepts in regards to the displayed body region and the directionality.
\end{table}

\section*{Limitations}
\label{sec:limitations}

The entire dataset is sourced from the Open Access Subset of the PMC database. This naturally introduces a bias in terms of the selected images on the one hand, and quality issues inherent to the PMC on the other hand.
One example of such quality issues is a lower image quality in the PMC archive compared to the published article. Another very rare, but often impossible to manually correct issue is the occasional mix-up of images, where images are reproduced with wrong captions, sometimes taken from a different publication.

A fundamental limitation is represented by faulty or fuzzy original captions that serve as ground truth. This became apparent during the process of manual concept creation and evaluation, where the annotators and radiologist involved reported various discrepancies. Affected captions involved, e.g., modality confusion where obvious CT images were labeled as MRI images, or ambiguous statements where pure CT images were labeled as PET/CT images because they were captured with a combined scanner unit. Another common issue was a lack of detail and context in the original captions. This included, for example, not specifying the modality or, in the case of follow-up images in a series, generally no context on the region depicted. Thus, samples may lack a sufficient set of concepts that would be needed to comprehend the contents and context of an image.

Due to the inherent imbalance in modality distribution, certain modalities such as positron emission tomography or combined modalities as per ROCO definition are relatively rare. Although this distribution reflects the rarity of these modalities within publications in general, it should be taken into account if the dataset is to be used in the context of rare modalities.

Semantic types representing concepts identifiable from images were selected based on consent and best effort, but not by dedicated medical personnel, due to the lack of a well-defined process and resources.

Additionally, to maximize sample size and variety, the dataset includes images with at least one concept. However, this may be a limiting factor, as single-concept images may lack the complexity required for comprehensive model training. Therefore, users are advised to consider the impact of single-concept images on the effectiveness of their models and adjust their selection criteria accordingly.

Several limitations apply in regards to the manually created concepts for image modalities, body regions (X-ray only), and projection directionalities (X-ray only), meant to validate automatically generated ones and to substitute missing ones.
The performed validation by a radiologist showed a generally high agreement in regards to modality- and (X-ray only) body region-related concepts.
In regards to the modality, only for the positron emission (DRPE) and the angiography (DRAN) modalities an agreement in the moderate respectively substantial range was observed. For the positron emission modality this be explained with the low sample count $n=2$) in the evaluation that fosters a strong bias in occasional disagreement. For the angiography modality, images labeled as X-ray modality by the radiologist can be ascribed to a conservative stance, e.g., barely visible traces of contrast agent without explicit mention of a performed angiography within the caption may have been labeled as X-ray modality.
In regards to the (X-ray only) body region, only for the pelvis and spine region an agreement in the moderate respectively substantial range was observed. For both, this can be explained by limitations in the determination of anatomical regions. For example, diseases such as osteoarthritis or advanced osteonecrosis of the hip might affect both the acetabulum and femoral head and thus might be categorized as either pelvis or lower extremity, leading to potential labeling inconsistencies. These will be resolved as part of future work through implementation of multi-label assignment. Additionally, some anatomical regions, such as those focused on the soft tissue of the neck, could not be assigned a class at all due to minor limitations within the IRMA classification system.
Yet, in regards to (X-ray only) directionality-related concepts the achieved moderate agreement can not only be explained by said conservative stance of the radiologist during evaluation. It further indicates the complexity of the given task, as even for experienced professionals distinguishing between anteroposterior and posteroanterior directionalities is non-trivial when not provided additional context. A general problem further lies in the very common lack of said additional context as well as the deliberate reduction of directional complexity to only four classes that occasionally do not leave room for sufficient differentiation. For instance, while labeling the directionality of a dorsopalmar hand radiograph as coronal posteroanterior is not entirely accurate, this was done here to not leave a notable amount of X-ray samples without a directionality label.
Due to these limitations, manual concepts have been documented distinctively, so dataset users have the possibility to decide on their own whether to use or exclude them from given concepts for images.

\section*{Usage Notes}

The images are provided exactly as they appear in the PMC Open Access Subset archive and must be resized or cropped before being used in a machine learning workflow. Also note that the images provided in the PMC Open Access Subset may be of different quality than the images included in the journal.

Some possible use cases for the ROCOv2~\cite{zenodo} dataset include pre-training models for handling radiological images, building systems to support structured medical reporting, as well as building multi-label medical concept classification models and caption prediction models as done in the ImageCLEFmedical Caption tasks, which can be used to support structured medical reporting. Another use case is the evaluation of deep learning models for multi-task learning.

Please see the GitLab repository mentioned in the next section for example scripts regarding baseline models and evaluation.

\section*{Code availability}

The code is available in a GitLab repository (available at \url{https://gitlab.com/saviola/rocov2-code}, accessed 2023-11-10).
The folder ``roco-2018'' contains scripts and models for the compound figure and radiological figure detection, as taken from the original ROCO pipeline, which are used to filter all extracted images of the PMC Open Access Subset for non-compound radiological images.

The folder ``baseline'' contains code for the training of the baseline models for concept detection and caption prediction, which is described in the Technical Validation section.

The folder ``ImageCLEF'' contains the pre-processing and evaluation scripts for the ImageCLEFmedical Caption 2023 challenge tasks.


\section*{Acknowledgements}
The authors thank Seyedeh Delaram Mirazziroudsari, Department of Computer Science, University of Applied Sciences and Arts Dortmund, Dortmund, Germany, for her support on X-ray directionality concept implementation. The work of Louise Bloch, Raphael Brüngel, Sven Koitka, and Obioma Pelka was partially funded by a PhD grant from University of Applied Sciences and Arts Dortmund, Dortmund, Germany. The work of Ahmad Idrissi-Yaghir and Henning Schäfer was funded by a PhD grant from the DFG Research Training Group 2535 Knowledge- and data-based personalisation of medicine at the point of care (WisPerMed).

\section*{Author contributions statement}

S.K. and O.P. conceived the models and workflows for creating the original ROCO dataset.
R.B., A.I., C.S., L.B., and J.R. wrote the original draft of the manuscript.
R.B., A.I., and H.S. performed formal dataset analysis and visualization.
A.I., S.K., O.P., J.R., and H.S. developed the software for the dataset creation workflow and evaluation.
L.B. and R.B. curated the dataset concept labels.
C.S. validated the dataset caption and concept quality.
A.A., C.F., A.H., P.H., H.M., and F.N. provided supervision.
All authors reviewed the manuscript.

\section*{Competing interests}

The authors declare no competing interests.

\renewcommand\thefigure{S\arabic{figure}}
\renewcommand\thetable{S\arabic{table}}
\setcounter{figure}{0}
\setcounter{table}{0}

\section*{Supplementary Information}

\begin{table}[h]
\centering
\caption{Frequency Analysis of Publication Types provided by PubMed. The table presents the publication type distribution of distinct PMC articles from which image/caption pairs have been extracted. Each article may be associated with multiple publication types. Note that the analysis is limited to those articles where publication types were explicitly available within the MEDLINE PubMed Citation database. 187 PMC articles were excluded due to not having standardized publication types available. Single publication type occurrences are omitted.}
\begin{adjustbox}{max height=\textheight, keepaspectratio}
\begin{tabular}{lr}
\toprule
\textbf{Publication Type} & \textbf{Frequency} \\
\midrule
Journal Article & 26,757 \\
Case Reports & 26,474 \\
Review & 2991 \\
Research Support, Non-U.S. Gov't & 1546 \\
Comparative Study & 293 \\
Observational Study & 193 \\
Letter & 178 \\
Evaluation Study & 152 \\
Randomized Controlled Trial & 128 \\
Research Support, N.I.H., Extramural & 122 \\
Clinical Trial & 116 \\
Multicenter Study & 98 \\
Systematic Review & 65 \\
Video-Audio Media & 39 \\
Retracted Publication & 38 \\
Validation Study & 32 \\
Editorial & 30 \\
Comment & 26 \\
Research Support, U.S. Gov't, Non-P.H.S. & 23 \\
Historical Article & 17 \\
Clinical Study & 14 \\
Meta-Analysis & 12 \\
Clinical Trial, Phase II & 11 \\
Clinical Trial Protocol & 10 \\
Practice Guideline & 9 \\
Controlled Clinical Trial & 9 \\
Research Support, N.I.H., Intramural & 8 \\
Research Support, U.S. Gov't, P.H.S. & 6 \\
Randomized Controlled Trial, Veterinary & 5 \\
English Abstract & 5 \\
Clinical Trial, Phase I & 4 \\
Published Erratum & 4 \\
Observational Study, Veterinary & 4 \\
Dataset & 4 \\
Clinical Conference & 3 \\
Congress & 2 \\
Technical Report & 2 \\
Clinical Trial, Phase III & 2 \\
Guideline & 2 \\
Equivalence Trial & 2 \\
\midrule
\textbf{Total} & \textbf{59,446} \\
\bottomrule
\end{tabular}
\end{adjustbox}
\label{tab:pub_type_freq}
\end{table}

\begin{figure}[H]
    \centering
    \includegraphics[width=\textwidth]{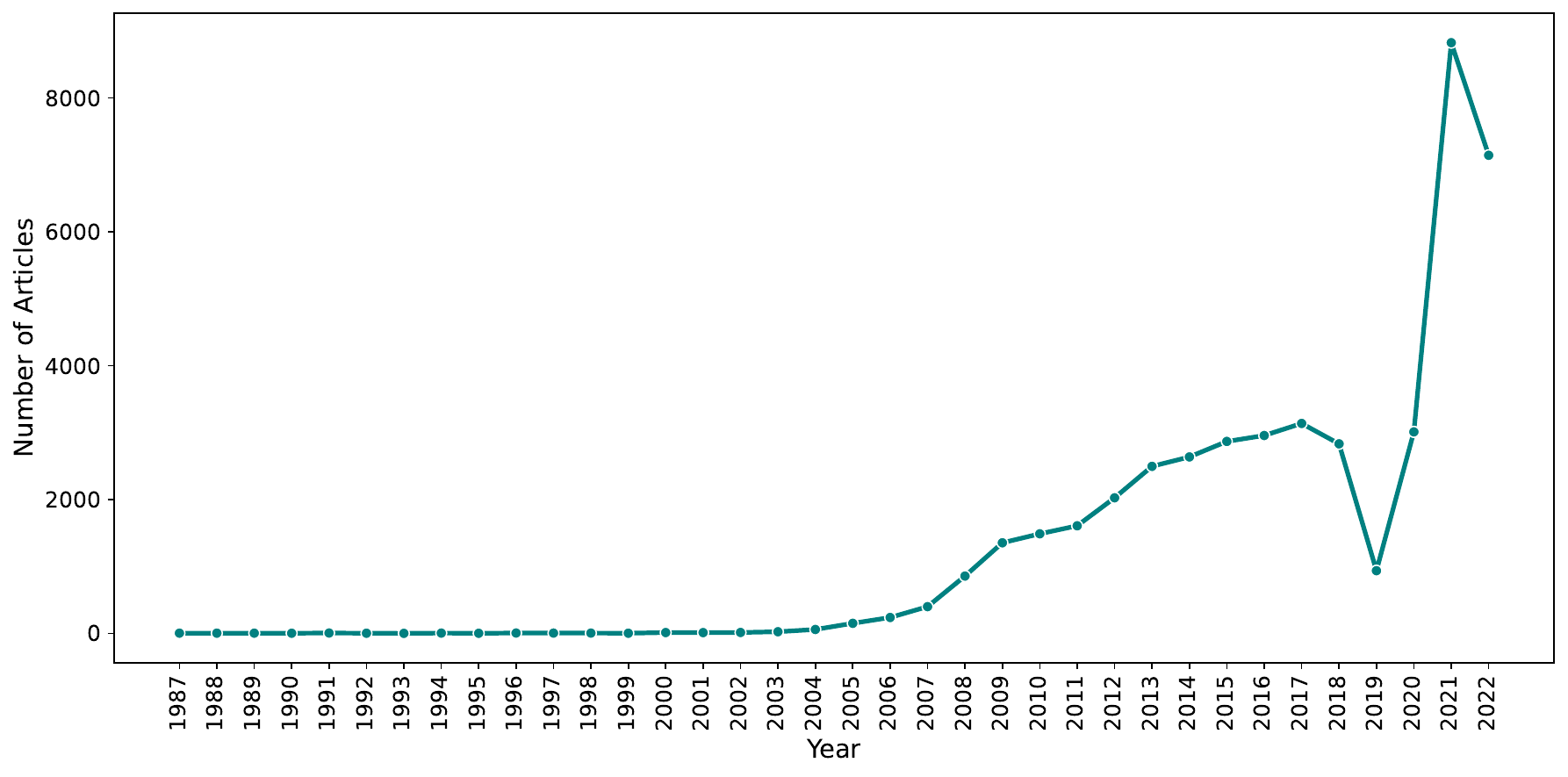}
    \caption{The distribution of article publications from 1987 to 2022, which were utilized for our data extraction. From 1987 to 2002, publications remained below 15 annually. A remarkable increase commenced in 2003, peaking in 2021 with 9341 articles. A dip in 2019 can be explained by regular updates to the dataset, which was originally released in 2018, beginning in 2020.}
    \label{fig:timeplot}
\end{figure}

\begin{figure}[H]
    \centering
    \includegraphics[width=\textwidth]{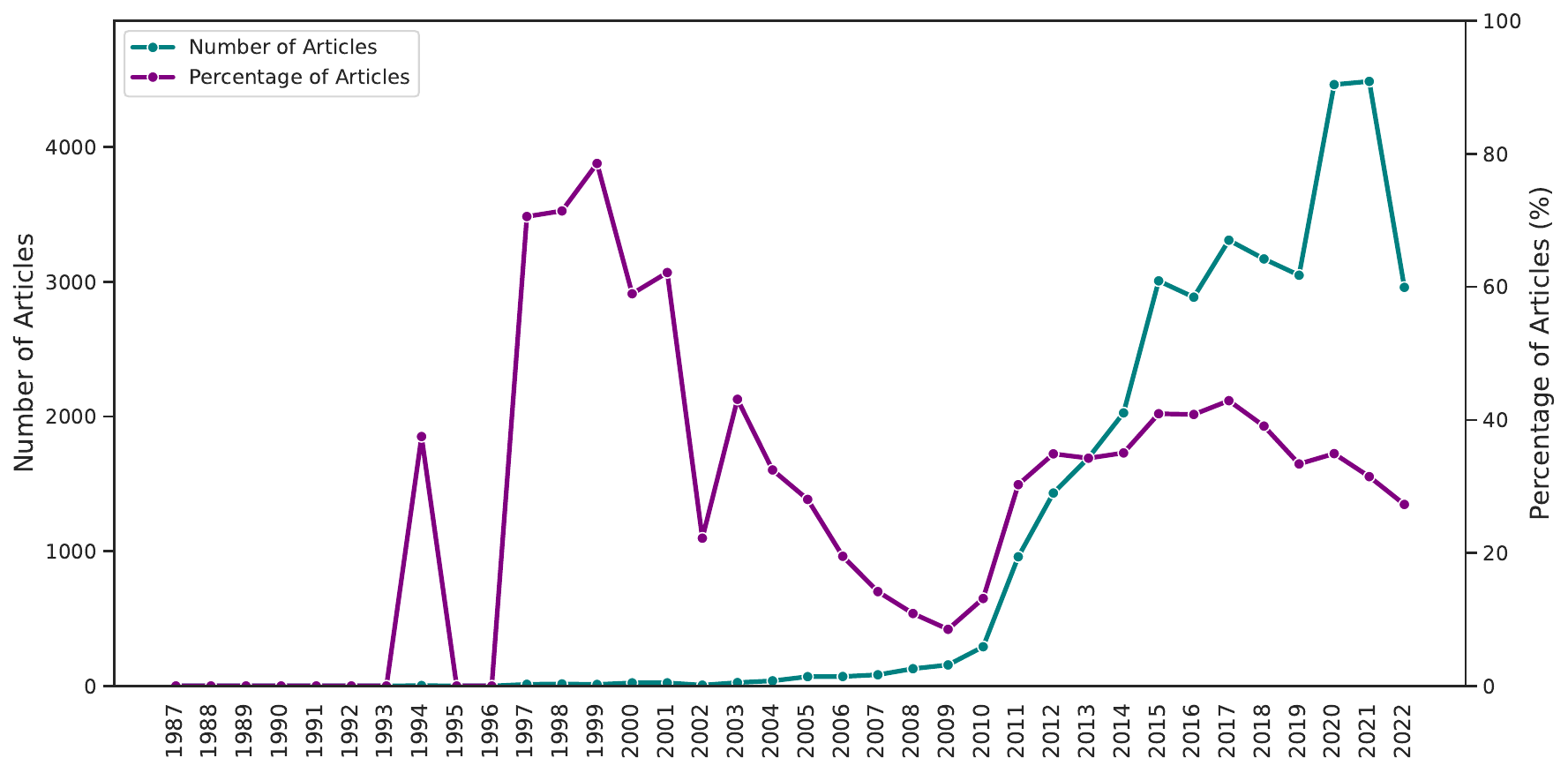}
    \caption{The relative and absolute distribution of CC BY-ND or CC BY-SA licensed PubMed articles from the years 1987 to 2022. Images from these articles were not considered for the dataset based on their license. When comparing the numbers with Figure~\ref{fig:timeplot}, it can be seen that these licenses have become increasingly popular from 2011 with a peak in 2017.}
    \label{fig:timeplot_removed}
\end{figure}

\begin{table}[ht]
\centering
\caption{Modality distribution of image and caption pairs within ROCOv2: Angiography modality (\texttt{DRAN}), combined modality (\texttt{DRCO}), CT modality (\texttt{DRCT}), MRI modality (\texttt{DRMR}), PET modality (\texttt{DRPE}), ultrasound modality (\texttt{DRUS}), and X-ray modality (\texttt{DRXR}).}
\begin{tabular}{lr}
\toprule
\textbf{Modality} & \textbf{Count} \\
\midrule
\texttt{DRCT} & 27,747 \\
\texttt{DRXR} & 21,997 \\
\texttt{DRMR} & 12,657 \\
\texttt{DRUS} & 11,429 \\
\texttt{DRAN} & 4799 \\
\texttt{DRCO} & 728 \\
\texttt{DRPE} & 432 \\
\midrule
\textbf{Total} & \textbf{79,789} \\
\bottomrule
\end{tabular}
\label{table:modality_distribution}
\end{table}

\begin{table}[h]
\centering
\caption{Distribution of manually created body part labels for the X-ray modality (\texttt{DRXR}).}
\begin{tabular}{lr}
\toprule
\textbf{IRMA body part} & \textbf{Count} \\
\midrule
\texttt{chest} & 7388 \\
\texttt{cranium} & 3973 \\
\texttt{lower\_extremity} & 3167 \\
\texttt{abdomen} & 2301 \\
\texttt{pelvis} & 1573 \\
\texttt{upper\_extremity} & 1478 \\
\texttt{spine} & 1363 \\
\texttt{breast} & 102 \\
\midrule
\textbf{Total} & \textbf{21,345} \\
\bottomrule
\end{tabular}
\end{table}

\begin{table}[h]
\centering
\caption{Distribution of manually created directionality labels for the X-ray modality (\texttt{DRXR}).}
\begin{tabular}{lr}
\toprule
\textbf{Category} & \textbf{Count} \\
\midrule
\texttt{coronal\_ap} & 9605 \\
\texttt{coronal\_pa} & 4303 \\
\texttt{sagittal} & 2425 \\
\texttt{transversal} & 51 \\
\midrule
\textbf{Total} & \textbf{16,384} \\
\bottomrule
\end{tabular}
\label{tab:my_label}
\end{table}

\begin{table}[h]
\caption{Top 100 journal distribution of PMC articles used in ROCOv2.}
\centering
\resizebox{\textwidth}{!}{
\begin{tabular}{lr|lr}
\toprule
\textbf{Journal} & \textbf{PMC Articles} & \textbf{Journal} & \textbf{PMC Articles} \\
\midrule
Cureus & 6249	&	Frontiers in Pediatrics & 164	\\
Journal of Medical Case Reports & 1514	&	International Journal of Clinical Pediatric Dentistry & 163	\\
The Pan African Medical Journal & 1396	&	Sage Open Medical Case Reports & 153	\\
Journal of Surgical Case Reports & 1128	&	Yonsei Medical Journal & 151	\\
Medicine & 782	&	Case Reports in Oncological Medicine & 149	\\
Clinical Case Reports & 773	&	Journal of Korean Neurosurgical Society & 146	\\
Cases Journal & 643	&	Case Reports in Pulmonology & 145	\\
Case Reports in Medicine & 597	&	The Journal of International Medical Research & 144	\\
Case Reports in Dentistry & 478	&	Imaging Science in Dentistry & 141	\\
World Journal of Surgical Oncology & 473	&	BMC Research Notes & 138	\\
PLoS One & 462	&	BMC Infectious Diseases & 137	\\
Case Reports in Surgery & 451	&	Case Reports in Emergency Medicine & 137	\\
Case Reports in Orthopedics & 426	&	Frontiers in Oncology & 137	\\
International Journal of Surgery Case Reports & 383	&	Journal of Medicine and Life & 134	\\
Journal of Cardiothoracic Surgery & 371	&	Frontiers in Surgery & 134	\\
World Journal of Clinical Cases & 349	&	Journal of Orthopaedic Case Reports & 130	\\
Journal of Clinical Medicine & 331	&	Journal of Community Hospital Internal Medicine Perspectives & 129	\\
Case Reports in Cardiology & 317	&	The Korean Journal of Thoracic and Cardiovascular Surgery & 127	\\
Clinical Practice and Cases in Emergency Medicine & 306	&	The Indian Journal of Radiology \& Imaging & 126	\\
Case Reports in Obstetrics and Gynecology & 301	&	Journal of the Belgian Society of Radiology & 126	\\
BMC Musculoskeletal Disorders & 299	&	Korean Circulation Journal & 125	\\
European Heart Journal. Case Reports & 295	&	Archives of Plastic Surgery & 124	\\
Diagnostics & 294	&	World Journal of Emergency Surgery & 121	\\
BJR Case Reports & 289	&	Iranian Journal of Radiology & 121	\\
Annals of Medicine and Surgery & 257	&	International Journal of Environmental Research and Public Health & 117	\\
Surgical Case Reports & 249	&	Clinics in Orthopedic Surgery & 116	\\
Case Reports in Urology & 242	&	BMC Gastroenterology & 115	\\
Korean Journal of Radiology & 230	&	Insights into Imaging & 113	\\
Scientific Reports & 230	&	Cardiovascular Ultrasound & 110	\\
Case Reports in Otolaryngology & 221	&	Case Reports in Critical Care & 110	\\
The Western Journal of Emergency Medicine & 220	&	Case Reports in Endocrinology & 107	\\
Case Reports in Gastroenterology & 219	&	Cardiology Research & 107	\\
Case Reports in Pediatrics & 212	&	Radiology Case Reports & 106	\\
Oxford Medical Case Reports & 211	&	Case Reports in Vascular Medicine & 105	\\
Korean Journal of Anesthesiology & 210	&	Clinical Endoscopy & 103	\\
Journal of Korean Medical Science & 206	&	Children & 101	\\
BMJ Case Reports & 205	&	F1000Research & 100	\\
Case Reports in Infectious Diseases & 202	&	Case Reports in Rheumatology & 98	\\
The Korean Journal of Internal Medicine & 196	&	Frontiers in Medicine & 97	\\
Case Reports in Oncology & 196	&	Journal of the Korean Association of Oral and Maxillofacial Surgeons & 95	\\
Journal of Orthopaedic Surgery and Research & 190	&	Journal of Clinical Medicine Research & 94	\\
Case Reports in Gastrointestinal Medicine & 188	&	Frontiers in Neurology & 94	\\
Medicina & 185	&	Gastroenterology Research & 91	\\
BMC Surgery & 183	&	BMC Cardiovascular Disorders & 90	\\
Radiologia Brasileira & 183	&	Case Reports in Neurological Medicine & 90	\\
Case Reports in Radiology & 179	&	Cancers & 90	\\
Journal of Investigative Medicine High Impact Case Reports & 176	&	Journal of Clinical and Experimental Dentistry & 90	\\
BioMed Research International & 170	&	Dental Press Journal of Orthodontics & 90	\\
Asian Spine Journal & 166	&	Diagnostic Pathology & 88	\\
Oncology Letters & 165	&	Case Reports in Pathology & 88	\\
\bottomrule
\end{tabular}
}
\label{tab:journals}
\end{table}

\begin{table}[ht]
\centering
\caption{Top 10 frequent CUIs for the angiography modality (\texttt{DRAN}).}
\begin{tabular}{llr}
\toprule
\textbf{CUI} & \textbf{UMLS Term} & \textbf{Images} \\
\midrule
C0002978 & Angiogram & 4799 \\
C0226032 & Anterior Descending Branch Of Left Coronary Artery & 516 \\
C0038257 & Stent, Device & 399 \\
C1261287 & Stenosis & 345 \\
C1261316 & Right Coronary Artery Structure & 338 \\
C0034052 & Pulmonary Artery Structure & 296 \\
C1947917 & Occluded & 257 \\
C0085590 & Catheter Device & 245 \\
C0001168 & Complete Obstruction & 220 \\
C0002940 & Aneurysm & 212 \\
\bottomrule
\end{tabular}
\end{table}

\begin{table}[ht]
\centering
\caption{Top 10 frequent CUIs for the combined modality (\texttt{DRCO}).}
\begin{tabular}{llr}
\toprule
\textbf{CUI} & \textbf{UMLS Term} & \textbf{Images} \\
\midrule
C0032743 & Positron-Emission Tomography & 232 \\
C1699633 & Pet/Ct Scan & 208 \\
C0034606 & Radionuclide Imaging & 74 \\
C0027651 & Neoplasms & 56 \\
C0040405 & X-Ray Computed Tomography & 51 \\
C0024204 & Lymph Nodes & 50 \\
C0028259 & Nodule & 36 \\
C0036525 & Metastatic To & 30 \\
C0023884 & Liver & 30 \\
C2939419 & Secondary Neoplasm & 27 \\
\bottomrule
\end{tabular}
\end{table}

\begin{table}[ht]
\centering
\caption{Top 10 frequent CUIs for the CT modality (\texttt{DRCT}).}
\begin{tabular}{llr}
\toprule
\textbf{CUI} & \textbf{UMLS Term} & \textbf{Images} \\
\midrule
C0040405 & X-Ray Computed Tomography & 27,747 \\
C0817096 & Chest & 2431 \\
C0000726 & Abdomen & 1651 \\
C0030797 & Pelvis & 1327 \\
C0444611 & Fluid Behavior & 838 \\
C0027651 & Neoplasms & 832 \\
C0023884 & Liver & 715 \\
C0205207 & Cystic & 661 \\
C0028259 & Nodule & 595 \\
C0225317 & Soft Tissue & 552 \\
\bottomrule
\end{tabular}
\end{table}

\begin{table}[ht]
\centering
\caption{Top 10 frequent CUIs for the MRI modality (\texttt{DRMR}).}
\begin{tabular}{llr}
\toprule
\textbf{CUI} & \textbf{UMLS Term} & \textbf{Images} \\
\midrule
C0024485 & Magnetic Resonance Imaging & 12,657 \\
C0006104 & Brain & 724 \\
C0027651 & Neoplasms & 486 \\
C0013604 & Edema & 466 \\
C0444611 & Fluid Behavior & 440 \\
C0037925 & Spinal Cord & 357 \\
C0205207 & Cystic & 332 \\
C0030797 & Pelvis & 304 \\
C0152295 & Cerebral White Matter Structure & 213 \\
C0018787 & Heart & 211 \\
\bottomrule
\end{tabular}
\end{table}

\begin{table}[ht]
\centering
\caption{Top 10 frequent CUIs for the PET modality (\texttt{DRPE}).}
\begin{tabular}{llr}
\toprule
\textbf{CUI} & \textbf{UMLS Term} & \textbf{Images} \\
\midrule
C0032743 & Positron-Emission Tomography & 432 \\
C0024204 & Lymph Nodes & 44 \\
C0025066 & Mediastinum & 30 \\
C0023884 & Liver & 27 \\
C0027651 & Neoplasms & 25 \\
C1266909 & Entire Bony Skeleton & 24 \\
C0037993 & Spleen & 20 \\
C0006826 & Malignant Neoplasms & 18 \\
C0036525 & Metastatic To & 18 \\
C0030797 & Pelvis & 18 \\
\bottomrule
\end{tabular}
\end{table}

\begin{table}[ht]
\centering
\caption{Top 10 frequent CUIs for the ultrasound modality modality (\texttt{DRUS}).}
\begin{tabular}{llr}
\toprule
\textbf{CUI} & \textbf{UMLS Term} & \textbf{Images} \\
\midrule
C0041618 & Ultrasonography & 11,429 \\
C0225897 & Left Ventricular Structure & 764 \\
C0225883 & Right Ventricular Structure & 611 \\
C0225860 & Left Atrial Structure & 429 \\
C0018827 & Heart Ventricle & 385 \\
C0205207 & Cystic & 384 \\
C0225844 & Right Atrial Structure & 361 \\
C0003483 & Aorta & 331 \\
C0018792 & Heart Atrium & 310 \\
C0031039 & Pericardial Effusion & 301 \\
\bottomrule
\end{tabular}
\end{table}

\begin{table}[ht]
\centering
\caption{Top 10 frequent CUIs for the X-ray modality (\texttt{DRXR}).}
\begin{tabular}{llr}
\toprule
\textbf{CUI} & \textbf{UMLS Term} & \textbf{Images} \\
\midrule
C1306645 & Plain X-Ray & 21,997 \\
C1999039 & Anterior-Posterior & 9605 \\
C0817096 & Chest & 7535 \\
C1996865 & Postero-Anterior & 4303 \\
C0037303 & Bone Structure Of Cranium & 3987 \\
C0023216 & Lower Extremity & 3183 \\
C0205129 & Sagittal & 2430 \\
C0000726 & Abdomen & 2375 \\
C0030797 & Pelvis & 1845 \\
C1140618 & Upper Extremity & 1478 \\
\bottomrule
\end{tabular}
\end{table}

\begin{table}[!htb]
\centering
\caption{Distribution of non-English captions.}
\label{tab:caption_distribution}
\begin{tabular}{lrr}
\toprule
\textbf{Language} & \textbf{Number of Captions} & \textbf{Proportion (\%)} \\ \midrule
French    & 1413 & 92.47 \\
Portuguese & 55   & 3.60 \\
Spanish   & 48    & 3.14 \\
Dutch     & 4     & 0.26 \\
German    & 4     & 0.26 \\
Italian   & 2     & 0.13 \\
Turkish   & 1     & 0.07 \\
Russian   & 1     & 0.07 \\ \midrule
\textbf{Total}     & \textbf{1528}  & \textbf{100.00} \\
\bottomrule
\end{tabular}
\end{table}

\begin{figure}[H]
    \centering
    \begin{subfigure}[t]{1\linewidth}
        \centering\includegraphics[width=1\linewidth]{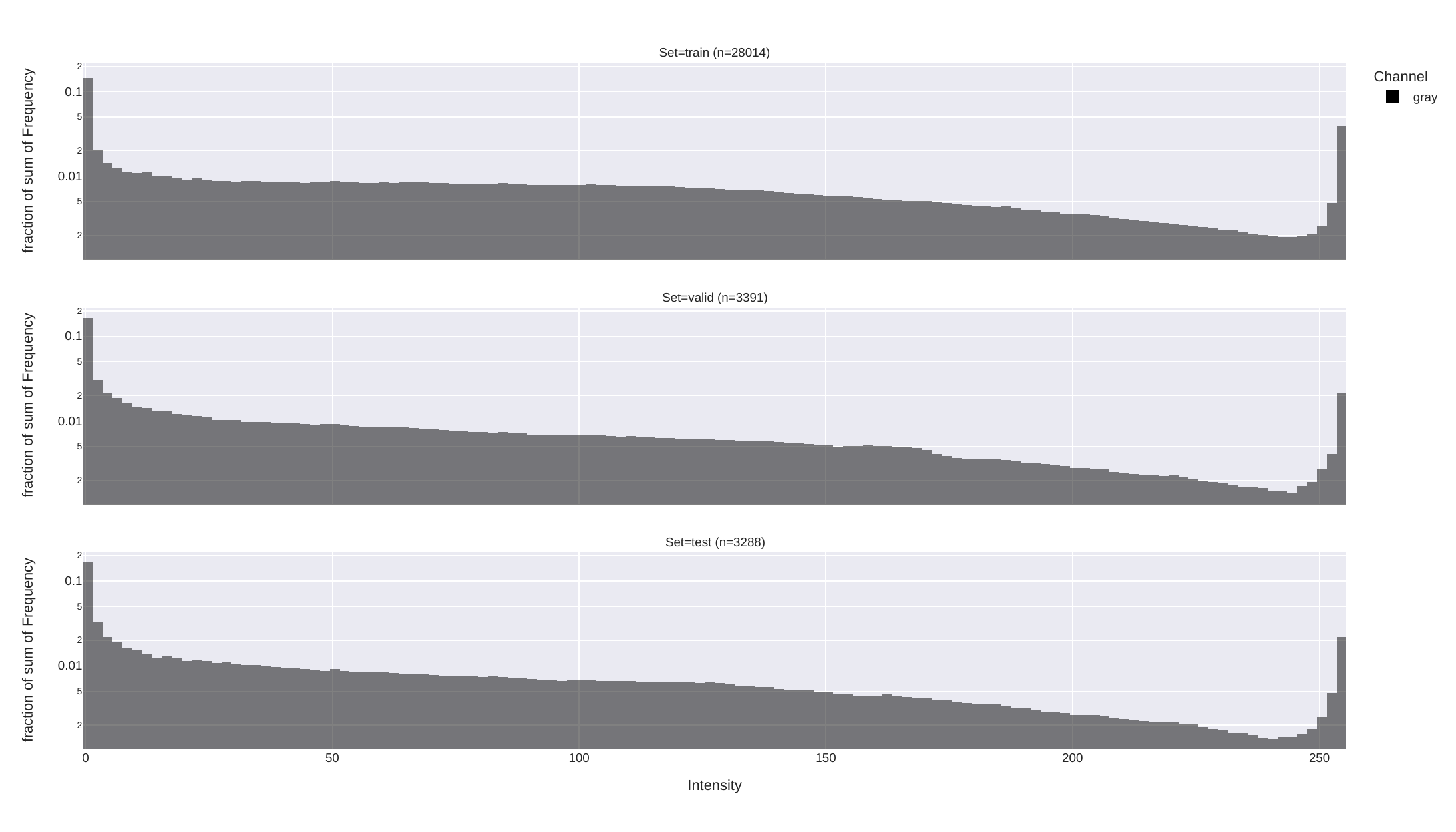}
        \caption{Grayscale images ($n=34,693$) intensity histograms (128 bins) of train, valid, and test sets.}
        \label{fig:channel_intensities_train}
    \end{subfigure}
    \begin{subfigure}[t]{1\linewidth}
        \centering\includegraphics[width=1\linewidth]{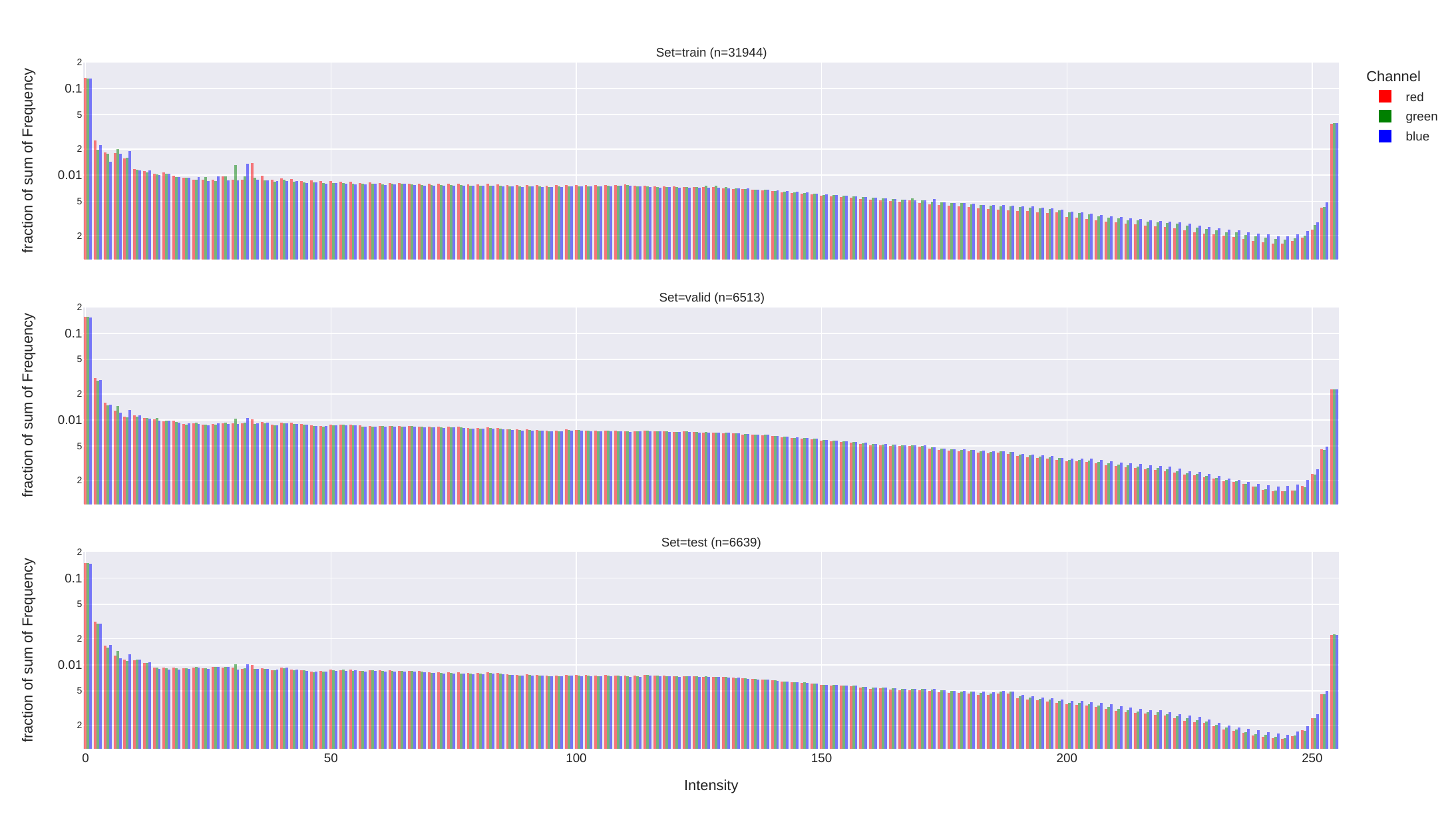}
        \caption{RGB images ($n=45,096$) intensity histograms (128 bins) of train, valid, and test sets.}
        \label{fig:channel_intensities_valid}
    \end{subfigure}
    \caption{Grayscale and RGB images ($n=79,789$) intensity histograms of train, valid, and test sets.}
    \label{fig:channel_intensities}
\end{figure}

\begin{figure}[H]
    \centering
    \begin{subfigure}[t]{1\linewidth}
        \centering\includegraphics[width=1\linewidth]{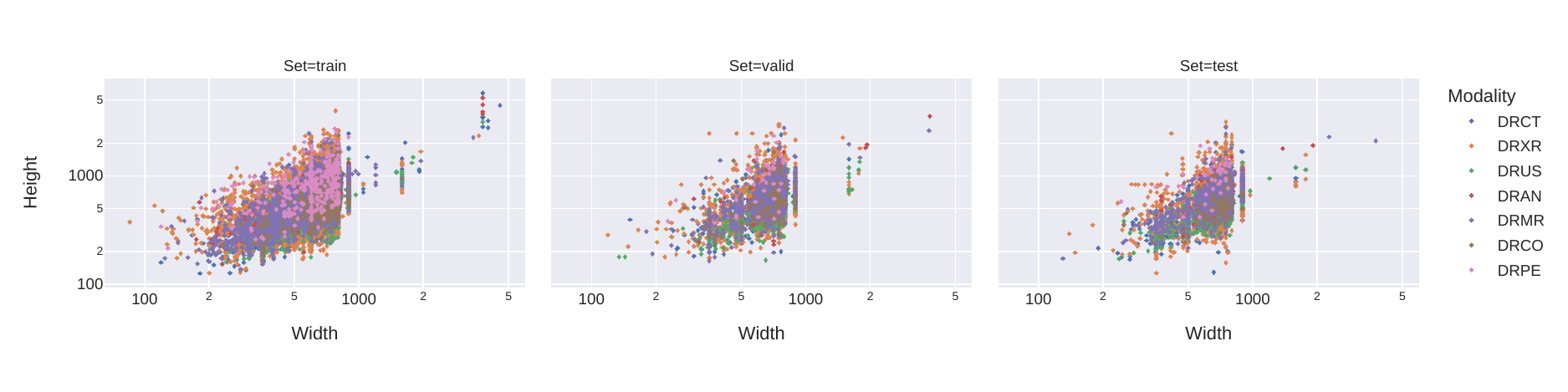}
        \caption{Image width, height distributions of train, valid, test sets.}
        \label{fig:width_height_images}
    \end{subfigure}
    \begin{subfigure}[t]{1\linewidth}
        \centering\includegraphics[width=1\linewidth]{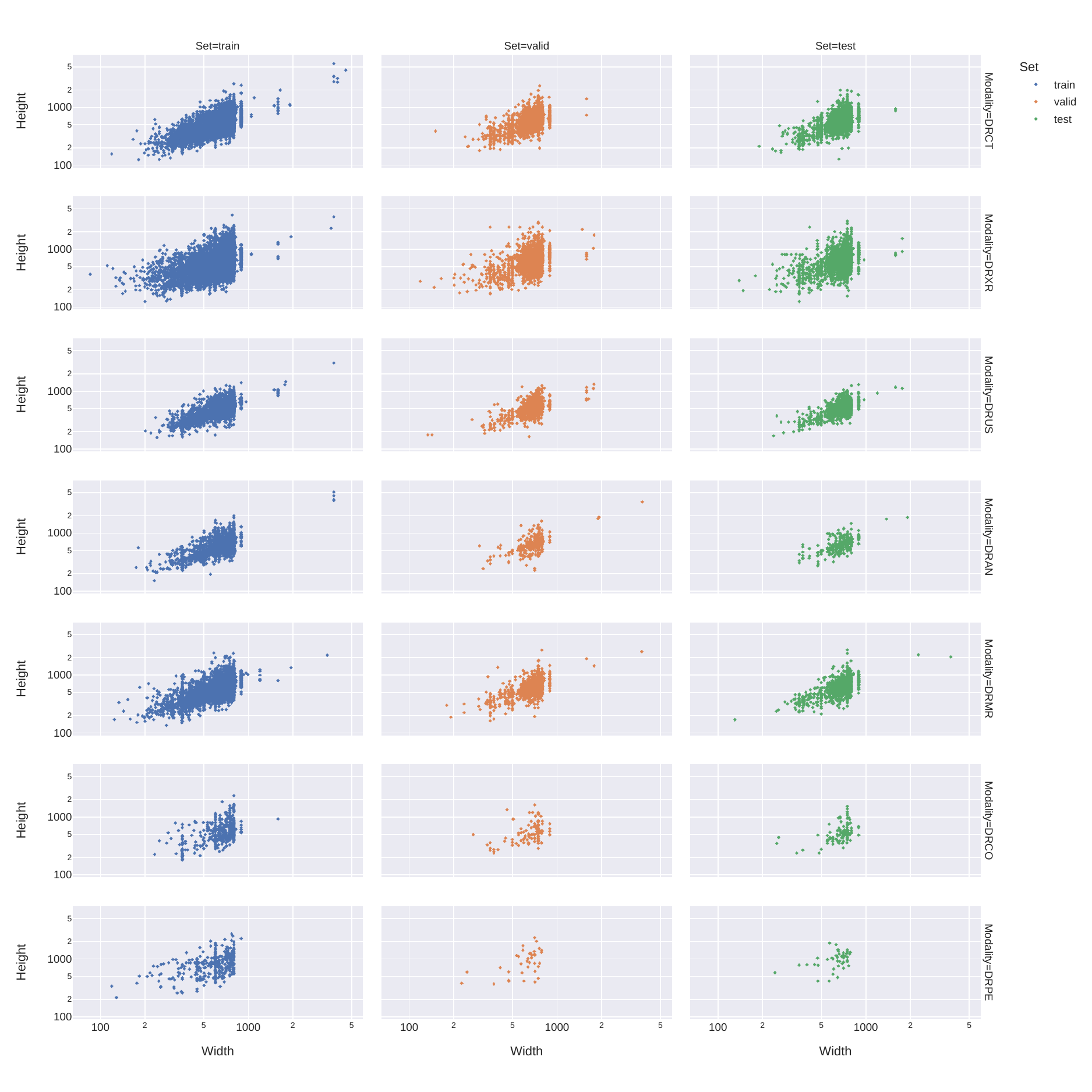}
        \caption{Image width, height distributions of train, valid, test sets separated by modality.}
        \label{fig:width_height_modalities}
    \end{subfigure}
    \caption{Image width, height distributions.}
    \label{fig:width_height_distributions}
\end{figure}

\begin{table}[ht]
\centering
\caption{Distilled annotation guideline for manual labeling of angiography (\texttt{DRAN}), combined (\texttt{DRCO}), CT (\texttt{DRCT}), MRI (\texttt{DRMR}), PET (\texttt{DRPE}), ultrasound (\texttt{DRUS}), and X-ray (\texttt{DRXR}) modalities.}
\label{tab:annotation_guideline_modality}
\begin{tabularx}{\textwidth}{l|X}
\toprule
\textbf{Label} & \textbf{Matching criteria} \\
\midrule
\multicolumn{1}{l}{General information:} & For each image the given caption is to be checked when its modality can not be determined with very high certainty by its visual features. In case its caption does not provide sufficient information or is ambiguous, checking its context within its original publication is demanded. Images may comprise annotations, e.g., arrows, markers, or sketches, up to a limit where these are dominating the visual. Images with a veterinary focus are to be included. \\
\midrule
\texttt{DRAN} & 2D X-ray projections that show application of contrast media for the specific purpose of highlighting blood vessels, e.g., ventriculography, aortography, or arterio/-venography. Angiographic methods performed using CT/MRI are not to be considered as part of this class, but as CT modality (\texttt{DRCT}) respectively MRI modality (\texttt{DRMR}). \\
\texttt{DRCO} & Combination of supported modalities, e.g., PET-CT/MRI, or SPECT-CT/MRI. However, CT/MRI scans with angiographic components are not considered. \\
\texttt{DRCT} & Classic CT scan images, e.g., single slices, or accumulations of few slices. Images involving contrast media are considered regular CT scans, also those with angiographic components. Vast 3D representations rendered from CT scan volume information are not considered. Combined with other imaging techniques, e.g. PET, SPECT, or likewise, images need to be considered as combined modality (\texttt{DRCO}). \\
\texttt{DRMR} & Classic MRI scan images, e.g., single slices, or accumulations of few slices. Images involving contrast media are considered regular MRI scans, also those with angiographic components. Vast 3D representations rendered from MRI scan volume information are not considered. Combined with other imaging techniques, e.g. PET, SPECT, or likewise, images need to be considered as combined modality (\texttt{DRCO}). \\
\texttt{DRPE} & Solely classic PET scans. Other similar but not positron emission-based methods, e.g., SPECT, or scintigraphy, are not considered. PET scans combined with CT/MRI are have to be considered as combined modality (\texttt{DRCO}). \\
\texttt{DRUS} & Classic 2D/3D ultrasound images. These may also represent user interface screenshots or comprise additional data on, e.g., flow series, Doppler visualization, or location. \\
\texttt{DRXR} & Classic 2D X-ray projections, but also X-ray panorama shots of the jaw region. The latter need to be differentiated from Cone Beam CT (CBCT) shots that share a certain resemblance. Methods using contrast media, e.g., in Barium swallow, urethrography, arthrogram, or cholangiography, are considered regular X-ray projections. However, methods using contrast media to specifically highlight blood vessels have to be considered as angiography modality (\texttt{DRAN}). \\
\midrule
\texttt{OTHER} & Label that pools diverse non-radiological and out-of-class images, later to be excluded from the raw dataset. This comprises compound images that incorrectly passed the compound filtering mechanism, vast 3D rendered images, synthetically generated images (e.g., by generative adversarial networks, or diffusion networks), photographs/sketches/collages (e.g., scenes, schematics, workflows, or multiple images of different contexts), non-medical modalities (e.g., transmission electron microscopy, or satellite images), non-supported modalities (e.g., optical coherence tomography (OCT), endoscopy, or histological slices), images representing a supported modality but not showing a specific medical context (e.g., assessment of materials, phantoms, or artifacts), images that contained captions as image parts and thus could leak information when using optical character recognition (OCR) techniques, and several others. \\
\texttt{UNKNOWN} & Label that pools images which can not be labeled with certainty due to insufficient information from visual features, captions, and publication context, later to be excluded from the raw dataset. \\
\bottomrule
\end{tabularx}
\end{table}

\begin{table}[ht]
\centering
\caption{Distilled annotation guideline for manual labeling of body regions for the X-ray modality (\texttt{DRXR}).}
\label{tab:annotation_guideline_irma}
\begin{tabularx}{\textwidth}{l|X}
\toprule
\textbf{Label} & \textbf{Matching criteria} \\
\midrule
\multicolumn{1}{l}{General information:} & For each image the given caption is to be checked when its body region can not be determined with very high certainty by its visual features. In case its caption does not provide sufficient information or is ambiguous, checking its context within its original publication is demanded. Images may comprise annotations, e.g., arrows, markers, or sketches, up to a limit where these are dominating the visual. The given context described by the caption is relevant, images showing multiple regions may not be excluded if the given context demands an expanded projection (e.g., a pronounced scoliosis displayed over chest and abdomen regions actually focuses the spine). The given projection directionality is not relevant. Images with a veterinary focus are to be excluded, as the underlying IRMA classification solely addresses human anatomy. \\
\midrule
\texttt{abdomen} & X-ray projections that cover a region ranging from the upper abdominal structures (e.g., diaphragma border, lower ribs, liver, stomach) down to the lower abdominal structures (e.g., bladder with urinary tract, gynecological organs, anus). A mild overlap with the chest region is acceptable, a moderate overlap may be acceptable given that the caption context addresses the abdomen region. A total overlap with the pelvis region is acceptable given the caption context clearly addresses the abdomen region. \\
\texttt{breast} & X-ray projections solely acquired from mammography. \\
\texttt{chest} & X-ray projections that cover a region from the upper thoracic structures (e.g., upper ribs, claviculae, upper pars thoracica of esophagus) down to the lower thoracic structures (e.g., thoracic diaphragma, lower ribs, lower lobes of lungs). A mild overlap with the cranium and abdomen regions is acceptable, a moderate overlap may be acceptable given that the caption context clearly addresses the chest region. Projections acquired via mammography are to be labeled as \texttt{breast}. \\
\texttt{cranium} & X-ray projections that mainly cover osseous cranial structures (e.g., cranium, mandibula, teeth). A mild overlap with the chest region is acceptable, a moderate overlap may be acceptable given that the caption context clearly addresses the cranium region. \\
\texttt{lower\_extremity} & X-ray projections that cover a region from hip joints (excluding the acetabulum as part of the pelvic bone) down to the toes. A mild overlap with the lower abdomen respectively pelvis regions is acceptable, a moderate overlap may be acceptable given that the caption context clearly addresses the lower extremity region. \\
\texttt{pelvis} & X-ray projections that mainly cover the osseous pelvic structures (e.g., pelvic bone, sacrum, acetabulum). A mild overlap with the lower extremity region is acceptable, a moderate overlap may be acceptable given that the caption context addresses the pelvic region. A total overlap with the lower abdominal region is unavoidable, yet the given caption context must clearly address the pelvic structures that are not covered by abdominal structures. \\
\texttt{spine} & X-ray projections that cover a region from C1 of the cervical spine down to L5 of the lumbar spine. A total overlap with the cranium, chest, abdomen, and pelvis region is unavoidable, yet the given caption context must clearly address the spinal structures. Contrary to other classes, full-body or full-torso projections that would be considered \texttt{OTHER} due to excessive region overlaps may be acceptable to be labeled as spine for cases that clearly address the whole spine (e.g., angle measurements, pronounced scoliosis, osteoporosis diagnostics). \\
\texttt{upper\_extremity} & X-ray projections that cover a region from glenohumeral joints down to the fingers. A mild overlap with the lower chest and abdomen regions is acceptable, a moderate overlap may be acceptable given that the caption context clearly addresses the upper extremity region. \\
\midrule
\texttt{OTHER} & Label that pools mixed-class and out-of-class images, that stay within the dataset but do not receive a manual annotations for body regions. This comprises images that either display a moderate to heavy overlap between regions without a clear focus on a specific region (e.g., too wide projection angles, cases multimorbidity presented in a single projection, pediatric full-body checks), veterinary cases, or very rare cases that can not be classified with certainty into the IRMA system (e.g., lateral projections of outer genitalia, lateral Barium swallow for upper esophagus cavity diagnostics).  \\
\texttt{UNKNOWN} & Label that pools images which can not be labeled with certainty due to insufficient information from visual features, captions, and publication context. These images stay within the dataset but do not receive a manual annotation. \\
\bottomrule
\end{tabularx}
\end{table}

\begin{table}[ht]
\centering
\caption{Distilled annotation guideline for manual labeling of directionality for the X-ray modality (\texttt{DRXR}).}
\label{tab:annotation_guideline_directionality}
\begin{tabularx}{\textwidth}{l|X}
\toprule
\textbf{Label} & \textbf{Matching criteria} \\
\midrule
\multicolumn{1}{l}{General information:} & For each image of the X-ray modality (\texttt{DRXR}), the directionality should be determined based on the standard exposures and projections. If the directionality is not clear from the image itself, additional context from the original publication or caption should be consulted. Images with a veterinary focus are to be excluded. \\
\midrule
\texttt{coronal\_ap} & Images that are taken in the coronal plane and are anteroposterior (AP) projections. This includes coronal chest X-rays taken with a portable device and in a supine or semi-erect position (e.g. emergency or bedside imaging) as well as standard coronal projections of the abdomen, spine, bones and joints of the pelvis, lower extremity, upper extremity (excluding hand and wrist) and, in some instances, the skull. \\
\texttt{coronal\_pa} & Images that are taken in the coronal plane and are posteroanterior (PA) projections. This includes coronal chest X-rays taken in a standing position, as well as standard coronal projections of the hand and wrist, skull, and, in some instances, the lumbar spine. \\
\texttt{sagittal} & Images that are taken in the sagittal plane, showing a lateral view of the body part. This includes lateral views of the chest, spine, bones and joints of the upper and lower extremities and, in some instances, abdomen. \\
\texttt{transversal} & Images that are taken in the transversal (axial) plane. This includes axial projections of the shoulder and the knee. \\
\midrule
\texttt{OTHER} & Label that pools mixed-class and out-of-class images, that stay within the dataset but do not receive a manual annotations for directionalities. This includes non-standard projections or views, such as dental X-rays (e.g., panoramic, bitewing, or periapical X-rays), special views of the skull (e.g., occipitofrontal, occipitomental, bregmaticooccipital, submentobregmatical, submentobregmaticofrontal, bregmaticooral, bregmaticosubmental), oblique angles (e.g., right anterior oblique (RAO), left anterior oblique (LAO)), or a combination of two projections within the same image. \\
\texttt{UNKNOWN} & Label that pools images which can not be labeled with certainty due to insufficient information from visual features, captions, and publication context. These images stay within the dataset but do not receive a manual annotation. \\
\bottomrule
\end{tabularx}
\end{table}

\end{document}